\documentclass{aa}
\usepackage{graphicx,amsmath,fixltx2e}
\usepackage{txfonts,color}

\usepackage{natbib} 
\bibpunct{(}{)}{;}{a}{}{,}

\begin{document}
\title{Thermodynamic fluctuations in solar photospheric three-dimensional
  convection simulations and observations} 

\author{C. Beck\inst{1,2,3} \and
  D. Fabbian\inst{1,2} \and F. Moreno-Insertis\inst{1,2} \and
  K.G. Puschmann\inst{4} \and R. Rezaei\inst{5}} 

\titlerunning{Thermodynamic fluctuations in photospheric convection simulations and observations}

\authorrunning{Beck et al.}  
\institute{Instituto de Astrof\'{\i}sica de
  Canarias (IAC)
  \and Departamento de Astrof{\'i}sica, Universidad de La Laguna (ULL)
   \and National Solar Observatory (NSO)
 \and Leibniz-Institut f\"ur Astrophysik Potsdam (AIP)
\and Kiepenheuer-Institut f\"ur Sonnenphysik   (KIS) } 
\date{Received xxx; accepted xxx} 

\abstract{Numerical three-dimensional (3D) radiative (magneto-)hydrodynamical [(M)HD] simulations of solar convection are nowadays used to understand the physical properties of the solar photosphere and convective envelope, and, in particular, to determine the Sun's photospheric chemical abundances. To validate this approach, it is important to check that no excessive thermodynamic fluctuations arise as a consequence of the partially incomplete treatment of radiative transfer causing too modest radiative damping.}
{We investigate the realism of the thermodynamics in recent state-of-the-art 3D convection simulations of the solar atmosphere carried out  with the \texttt{Stagger} code.}
{We compared the characteristic properties of several \ion{Fe}{i} lines
  (557.6\,nm, 630\,nm, 1565\,nm) and one \ion{Si}{i} line at 1082.7\,nm in
  solar disc centre observations at different spatial resolution with spectra
  synthesized from 3D convection simulations. The observations were taken with
  ground-based (Echelle spectrograph, G\"ottingen Fabry-P\'erot Interferometer (GFPI), POlarimetric LIttrow Spectrograph, Tenerife Infrared Polarimeter, all at the Vacuum Tower Telescope on Tenerife) and space-based instruments (Hinode/SPectropolarimeter). We degraded the synthetic spectra to the spatial resolution of
  the observations, based on the distribution of the continuum intensity
  $I_c$. We estimated the spectral degradation to be applied to the simulation results by comparing atlas spectra with averaged observed spectra. In addition to deriving a set of line parameters directly from the intensity profiles, we used the SIR (Stokes Inversion by Response functions) code to invert the spectra.}
{{\bf The spatial degradation kernels yield a generic spatial stray-light contamination between $\sim$20\,\% and $\sim$70\,\%.\rm} The spectral stray light
  inside the different spectrometers is found to be between 2 and 20\,\%. Most of the line parameters from the observational data are matched well by the degraded HD simulation spectra. The inversions
  predict a macroturbulent velocity $v_{\rm mac}$ below 10\,ms$^{-1}$ for the
  HD simulation spectra at full spatial resolution, whereas they yield $v_{\rm mac} \lesssim 1000$\,ms$^{-1}$ at a spatial resolution of 0\farcs3. The temperature
  fluctuations in the inversion of the degraded HD simulation spectra do not exceed those from the observational data (of the order of 100-200\,K rms for  $-2 \lessapprox \log \tau_{\rm 500nm} \lessapprox -0.5$). The comparison of line parameters in spatially
  averaged profiles with the averaged values of line parameters in spatially
  resolved profiles indicates a significant change of (average) line properties at a spatial scale between 0\farcs13 and 0\farcs3.}
{Up to a spatial resolution of 0\farcs3 (GFPI spectra), we find no indications of the presence of excessive thermodynamic fluctuations in the 3D HD simulation. To definitely confirm that simulations without spatial 
  degradation contain fully realistic thermodynamic fluctuations requires
  observations at even better spatial resolution (i.e.~$<$\,0\farcs13).}

\maketitle
\section{Introduction}
The abundances of chemical elements in the solar atmosphere are crucial as a
reference of chemical abundances throughout the universe, with a profound
influence on planetary science, stellar evolution theory, or the
interpretation of stellar spectra. Solar abundances are commonly derived from
comparisons between observed spectral lines and theoretical spectra from
atmospheric models with a varying degree of sophistication \citep[see][and
references therein]{asplund+etal2005a}. Such models range from static 1D
atmospheres treated in local thermal equilibrium (LTE) to dynamical
three-dimensional (3D) atmospheres with a non-LTE treatment \citep[e.g.][]{shchukina+etal2012}.

For static 1D solar atmosphere models, such as those of
\citet{holweger+mueller1974}, \citet{vernazza+etal1981},
\citet{fontenla_etal_06}, and variations thereof, one must first adopt values
for the solar abundance of the various chemical elements to derive a
temperature stratification through the comparison of observed and
synthetic spectra, using both the locations of bound-free edges, i.e.~the limiting wavelength of relevant ionization transitions, and the shape of
spectral lines. However, the retrieved temperature stratifications of the atmospheres are then used to derive the abundances by comparison
of up to a few tens of resulting synthetic spectral lines to
observations. This is therefore partly a case of circular reasoning, with the abundances both as input and output parameter. 

Another problem of static 1D models is that the dynamical nature
of the solar atmosphere cannot be taken into account consistently. Resolved
or unresolved velocities can only be partly included in the static 1D models
by using generic macroturbulent ($v_{\rm mac}$) and microturbulent ($v_{\rm
  mic}$) velocities. These are free parameters that show a certain
degree of degeneracy when one tries to disentangle different contributions to
the formation process of spectral lines \citep{allendeprieto+etal2001b}.

The accuracy of the information about dynamical variations provided by 
the observations is also always limited, since observed spectra
suffer from the combined effects of limited spatial, spectral, and temporal
resolution. The accurate retrieval of all thermodynamic parameters needed for an exact characterization of the dynamics, such as temperature, mass velocities, and mass densities, is therefore only partly possible on the observational side.

To obtain a realistic dynamical model of the solar atmosphere, one thus has
to resort to numerical simulations. The growing available computing power has
made it possible to run increasingly large-scale and higher-sampling
multi-dimensional calculations solving the complex set of coupled
differential equations necessary to model the relevant physics in the solar
photosphere
\citep[e.g.][]{nordlund+stein1990,steiner+etal1998,stein+nordlund1998}. When
one attempts, however, to extend such simulations to cover also the solar
chromosphere, non-LTE effects appear that cannot be (fully) included in 3D
numerical simulations that are of a sufficiently large scale while remaining
computationally feasible \citep[cf.][]{rammacher+cuntz1991,skartlien2000,asplund+etal2003,leenaarts+etal2007,leenaarts+etal2009,hayek+etal2010,wedemeyer+carlsson2011,carlsson+leenaarts2012,leenaarts+etal2012}.
Similarly, a possible problem of photospheric simulations is that radiative
transfer through the atmosphere can only be included in an approximate way,
commonly by using opacity binning
\citep[e.g.][]{nordlund1982,skartlien2000,voegler2004,voegler+etal2004}.

Nonetheless, the results of such numerical simulations are considered
realistic enough to be used in approaches to determine solar abundances
\citep[e.g.][]{asplund+etal2000}. The advantages of these simulations are
that they cover the dynamics of mass motions without the need to resort to
ad-hoc parameters such as $v_{\rm mac}$ or $v_{\rm mic}$. Furthermore, they
directly contain all information including especially the gas density on an absolute geometrical height scale, while,
e.g.~spectral line inversions just provide information of physical parameters on an optical depth scale. The
conversion to a geometrical height scale is in the latter case not
straightforward, needing sophisticated methods and certain assumptions commonly in the form of a boundary condition on electron (or gas) pressure or density 
\citep[e.g.][]{almeida2005,puschmann+etal2005,puschmann+etal2010,beck2011}.

However, the use of 3D HD simulations
coupled with improved atomic data and differences in equivalent width
measurements retrieved from observations resulted in the course of the last
decade in a downward revision of the solar abundance of mainly C, N and O
\citep[][]{asplund+etal2005a,asplund+etal2009}, which significantly
deteriorated the previous agreement between predicted and observed sound
speeds in the solar interior derived from helioseismology
\citep{bahcall+etal2005}. We note that, where available, non-LTE corrections
\citep[e.g.][]{fabbian+etal2009,fabbian+etal2009a} were included in the
abundance determination from simulations \citep[cf.][and references
  therein]{asplund+etal2009}. More recently, magnetic fields have been
suggested as another source of appreciable uncertainty
\citep{fabbian+etal2010,fabbian+etal2012}.

\citet{kalkofen2012} made an interesting comment about the use of simulation
results as a reference. The atomic abundances are usually derived by
matching the observed solar spectrum with the space- and/or
time-averaged synthetic line profiles from the simulations. In the averaging process, the
thermodynamic properties of the simulations are smeared out. Because of the
simplified treatment of the radiative transfer in the simulations, an
important damping factor could be missing. This could lead to
unrealistic/excessive thermodynamic fluctuations both in temperature and mass
motions. As a consequence of, e.g.~too large velocity fluctuations, the
average predicted profile would become too broad, thus leading to a 
  downward revision of the solar abundance reference of a given chemical
element, as derived by matching observational spectra. The argument by
\citet{kalkofen2012} is based on the fact that the intrinsic thermodynamic
parameters in simulations at full spatial resolution were investigated only
partially up to now.

The agreement of synthetic and observed data was verified in terms of the
contrast of continuum intensity, both at disc centre and for its
centre-to-limb (CLV) variation
\citep[e.g.][]{asplund+etal2000,wedemeyer+etal2009,asplund+etal2009,afram+etal2011},
as well as in terms of the match between average synthetic and observed
profiles \citep[e.g.][]{asplund+etal2000}. Spatially resolved profiles were
later also studied for comparison purposes \citep{pereira+etal2009}. All of
these comparisons, however, have a clear limitation: the
observations employed always have a lower spatial resolution than the original simulations,whose spatial grid size can be as small as several km. The good match with observations that was achieved in the aforementioned studies therefore proves that the simulations can be spatially and spectrally degraded to the properties of the observations, but it actually does not exclude
excessively large thermodynamic fluctuations in the simulations themselves,
as pointed out by \citet{danilovic+etal2008}.

In this paper, we investigate the realism of the thermodynamics in
numerical 3D convection simulations by comparing disc centre observations in
several spectral lines with the results from simulations. This shall serve as
the first step in a thorough derivation of the solar oxygen abundance using
the same set of simulations. In a subsequent study, we will turn to the CLV
of selected spectral lines and in the continuum, which can allow one to
calibrate the efficiency of H collisions in non-LTE calculations or to
constrain the temperature stratification.
\begin{table*}
\caption{Instruments and data properties, and corresponding settings of the spectral synthesis for the simulations. \label{table1}}
\begin{tabular}{c|cccccc|c}\hline\hline
instrument  & SP  & GFPI  & POLIS & TIP  & TIP   & Echelle     & Simulation\cr
$\lambda$ [nm] & 630 & 630  &  630 & 1083 & 1565 & 557   & all \cr
 main line(s)    & \ion{Fe}{i} & \ion{Fe}{i}  &
 \ion{Fe}{i}, \ion{O}{i}  & \ion{Si}{i} & \ion{Fe}{i}     & \ion{Fe}{i}, \ion{O}{i} &  \ion{Fe}{i}, \ion{Si}{i}\cr
dispersion [pm]  & 2.15 & 2.31  &  1.49  & 2.10 & 2.95  & 0.31  & 0.50 \cr
sampling [``]   & 0.3x0.3  & 0.11x0.11 & 0.5x0.15 & 0.5x0.18 & 0.5x0.18 & 0.18x0.18 & 0.13x0.13 \cr 
FOV [``]  &115x115 & 30x53 & 100x94  & 100x78 & 40x75  & 35x200  & 8.3x8.3\cr
integ. time [s]  & 1.6   & 0.64 &  3.3  & 3.3 & 4   & 3 & --\cr
spatial resolution\tablefootmark{1} [``]  & 0.64\tablefootmark{2} & 0.3/0.6\tablefootmark{3} & 0.7 & 0.7 & 0.7 & $\sim$1   & 0.13\tablefootmark{4} \cr
app. \# of profiles & 150.000 & 140.000 & 130.000 & 90.000 & 30.000 & 220.000 & 4.000\cr
\end{tabular}\\$ $\\
\tablefoottext{1}{estimated values, cf.~Sect.~\ref{spatdeg} below}
\tablefoottext{2}{sampling limited resolution}
\tablefoottext{3}{different data reduction methods, estimated spatial resolution values from PuB11}
\tablefoottext{4}{sampling $\equiv$ resolution}
\end{table*}

Section \ref{sec_obs} and \ref{sec_sim}, respectively, describe the
observations and the modeling employed here. The data reduction and the
analysis methods are explained in Sect.~\ref{sec_data}. Results are presented
in Sect.~\ref{sec_res} and are discussed in Sect.~\ref{sec_disc}. Section
\ref{sec_concl} provides our conclusions. The Appendices contain the
results for some of the spectral lines not discussed in the main text.

\section{Observations\label{sec_obs}}
We used several quiet Sun (QS) observations at disc centre taken with
different instruments and telescopes (see Table \ref{table1}). We
introduce them here divided into three groups:

\paragraph{(a) The infrared range.} For all spectral lines in the (near-)infrared (IR) range (\ion{Si}{i} 1082.7\,nm, \ion{Fe}{i} 1564.8\,nm, \ion{Fe}{i} 1565.2\,nm), the Tenerife Infrared Polarimeter \citep[TIP,][]{martinez+etal1999,collados+etal2007} at the German Vacuum Tower Telescope \citep[VTT,][]{schroeter+soltau+wiehr1985} was employed. The data in these lines were large-area scans. The
data at 1083\,nm were taken during a velocity calibration campaign of the
CHROmospheric TELescope \citep[CHROTEL,][]{bethge+etal2011} on 08 Dec 2007
between UT 10:02:35 and 10:17:55. The data at 1565\,nm were taken on 21 May
2008 from UT 08:44:18 until 08:51:43, in between the two long-integrated
data sets used in \citet{beck+rezaei2009}, and cover the same field of view
(FOV) as the data presented in the latter article. Spatial and spectral
sampling, total FOV, the integration time, and the approximate number of
profiles in the FOV are listed in Table \ref{table1}. The integration time
denotes the total exposure time that entered in a single measurement of the
full Stokes vector at one wavelength position or one slit position. 

\paragraph{(b) The 630 nm lines.}
The data in the \ion{Fe}{i} line at 630.25\,nm, taken with the G\"ottingen
Fabry-P\'erot Interferometer \citep[GFPI,][]{puschmann+etal2006,nazi+kneer2008a} at the VTT on 06 Jun 2009, are described in detail in \citet[][PuB11]{puschmann+beck2011}. We use the same spectral scan as those authors, with data reduced on the one hand with the multi-object multi-frame blind deconvolution
\citep[MOMFBD,][]{vannoort+etal2005} method, and on the other hand only
destretched to a reference image that was speckle-reconstructed with the method of \citet[][]{puschmann+sailer2006}. The spectral line was sampled on 28 wavelength steps. The data at 630\,nm from the POlarimetric LIttrow Spectrograph \citep[POLIS,][]{beck+etal2005a} were taken simultaneously to the TIP data in 1082.7\,nm on 08 Dec 2007. \citet{bethge+etal2012} discuss another example of data taken with this setup on the same day. The Hinode/SP \citep[][]{kosugi+etal2007,tsuneta+etal2008} map was taken during HOP 0190 on 05 Sep 2011 between UT 07:34:50 and 07:59:03. The FOV was in this case located slightly off disc centre at a heliocentric angle of less than 10 degrees, which should, however, not have a strong impact on the spectra. 

\paragraph{(c) The 557 nm range.} All the observations described so far provided
Stokes vector measurements that allow one to locate any magnetic flux inside
the FOV of sufficient strength and to determine the magnetic field
properties. Data from a wavelength range around 557\,nm were used in
addition. They correspond to a part of the CLV observations that we collected
as an ingredient to be used in the derivation of the solar oxygen abundance
\citep[][{\it in prep.}]{fabbian+etal2013}, as a reappraisal of previous
literature estimates
\citep[cf.][]{asplund+etal2004,melendez+asplund2008,pereira+etal2009,pereira+etal2009a}.
These spectra were obtained with the Echelle spectrograph of the VTT at high
spectral resolution, but without polarimetry. Additionally to an \ion{O}{i}
line at 557.7\,nm, the spectral range encompassed also the strong \ion{Fe}{i}
line at 557.6\,nm. We therefore included in the present study a disc centre
data set that we collected on 21 Nov 09 between UT 08:50 and 09:05.

All ground-based observations profited from the
real-time correction of seeing by the Kiepenheuer-Institut adaptive optics
(AO) system \citep[][]{vdluehe+etal2003} at the VTT.
\section{Convection simulations and spectral synthesis\label{sec_sim}}
The synthetic spectra used in our analysis are based on a set of 3D solar
convection simulations obtained and described by
\citet{fabbian+etal2010,fabbian+etal2012} that were carried out using
the \texttt{Stagger}\, code \citep[see,
  e.g.][]{stein+etal2011,kritsuk+etal2011,beeck+etal2012}. The simulation
box spanned approximately 6 x 6\,Mm$^2$ horizontally, covering $\sim$\,15
granules at any given time and with 252 equidistant points in each direction,
resulting in a horizontal grid spacing of $\sim 24$\,km. The height range was
covered by 126 grid points, reaching from $\sim 0.5$\,Mm above the height
corresponding to the optical depth unity ($\tau_{500\,\mathrm{nm}} = 1$)
level down to $\sim 2.0$\,Mm below it, with non-uniform vertical sampling
having a finest spacing of about 15\,km. 

The \texttt{Stagger}\, code uses a sixth-order finite differences
spatial scheme and a third-order Runge-Kutta temporal scheme to solve
the equations for the conservation of mass, momentum and energy.  At
the same time, the radiative transfer equation is solved assuming LTE,
in this instance along nine rays (one per octant, plus the vertical).
Four opacity bins were employed to model the wavelength dependency of
the radiation field. For more details on the opacity binning technique
adopted in the code see, e.g., \citet{nordlund1982},
\citet{collet+etal2011} and \citet{beeck+etal2012}.

The effective temperature in the simulation snapshots is controlled by
the (fixed) entropy density condition imposed on fluid flowing in from
the bottom boundary. This parameter was adjusted so that the resulting
T$_{eff}$ in the atmospheric models (about 5730\,K) was within less than one percent of the solar effective temperature derived from observations. \citet[][their
Fig.~2]{fabbian+etal2012} have shown that the continuum intensity computed 
for the same series of snapshots as employed here agrees reasonably well with literature data, albeit being slightly ($\lessapprox 5$\,\%) lower (cf.~also the bottom left panel in Fig.~\ref{av_prof} below). The variation of the emerging radiative flux at the surface for the different (M)HD cases corresponds to an uncertainty on T$_{eff}$ of the order of 0.2\,\% only, i.e.~about 10\,K. The simulation runs covered the field-free purely hydrodynamical (HD) case and three magneto-hydrodynamical (MHD) cases with a vertical unipolar magnetic flux density of 50, 100, and 200\,G, respectively.

In the present investigation we selected only the last snapshot of the
statistically stationary regime of each (M)HD run. The spectral synthesis was
performed using the LTE code LILIA \citep{socasnavarro2001}. For this step,
the appropriate input data corresponding to the chosen snapshot were prepared
by trimming the simulation results to a vertical extent ranging from $\sim\,+425$\,km above the average $\tau_{500\,\mathrm{nm}}\,=\,1$
level to $\sim\,-475$\,km below it, and by interpolating to a grid of equidistant points over the new physical depth. Those heights correspond to an optical
depth range from about $\log \tau_{500\,\mathrm{nm}}\,\approx\,-4$ to $\log
  \tau_{500\,\mathrm{nm}}\approx+2$, but with some variation across the
  FOV. The final input data cube has $115$ depth points with a constant, finer
  vertical grid spacing of $\sim 7.8$ km and $115$ depth points, which is
  better suited to our spectral synthesis purpose. 

Finally, for the spectral
  synthesis we also reduced the horizontal sampling to only include
  63$\times$63 grid points (without binning or averaging of the corresponding
  physical values) distributed over the same area as that of the original
  snapshot. Tests done for \citet{fabbian+etal2012} indicated that this
    approach does not introduce any kind of bias in the resulting set of
    synthetic spectra, while still providing a sufficiently large statistical
    base of about 4000 independent spectra. The final spatial sampling then
  corresponds to 0\farcs13, which is more than twice better than the best
  spatial resolution in the observations (cf.~Table \ref{table1}). 

The spectral sampling for all synthetic lines was set to 0.5\,pm. The synthetic spectra were computed using the Van der Waals broadening formula in the approximation of \citet{unsoeld1955}.  The default solar abundance adopted by LILIA was used, i.e.~the one given in \citet{grevesse1984}, according to which the logarithmic solar abundance of iron is log(Fe)=7.5~dex for the Sun, on the usual scale of 10$^{12}$ atoms of hydrogen, i.e.~log(H)=12.0~dex.

The spectral lines considered in the present study form in the low to mid 
  photosphere \citep[cf.][]{cabrera+bellot+iniesta2005}. All of them can be
  assumed to behave close to LTE except for the \ion{Si}{I} line at
  1082.7\,nm, where the line core is sensitive to non-LTE effects and to a lesser amount sensitive to 3D effects (e.g.~\citeauthor{wedemeyer2001}  \citeyear{wedemeyer2001};  \citeauthor{shchukina+trujillobueno2001}  \citeyear{shchukina+trujillobueno2001};  \citeauthor{shi_etal2008}  \citeyear{shi_etal2008}; \citeauthor{shchukina+etal2012}  \citeyear{shchukina+etal2012}; see also the corresponding sections in \citeauthor{asplund2005} \citeyear{asplund2005}). The line core of the \ion{Si}{i} line also forms in the uppermost layers of the simulation box \citep[up to 0.5\,Mm height;][]{bard+carlsson2008,felipe+etal2010} where the results of both the simulation and the spectral synthesis become less reliable.
\section{Data reduction and analysis methods\label{sec_data}}
\subsection{Intensity normalization}
\begin{figure}
\resizebox{8.8cm}{!}{\includegraphics{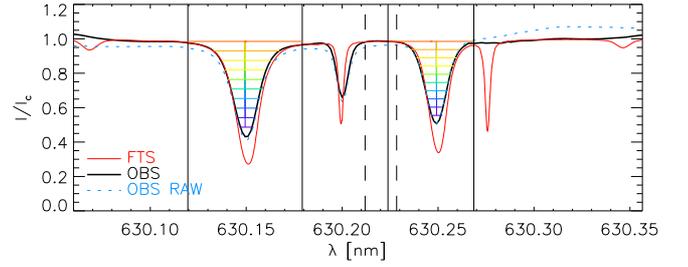}}
\caption{Example spectrum at 630\,nm from POLIS. Blue dotted line:
  average spectrum without correction for intensity gradients over the
  wavelength range. Black line: corrected spectrum. Red line: FTS atlas spectrum. The dashed vertical lines denote the ``continuum'' range, the solid vertical lines the range that was considered for determining the line parameters of \ion{Fe}{i} 630.15\,nm and  \ion{Fe}{i} 630.25\,nm, respectively. The colored horizontal lines inside the absorption profiles illustrate the method used to determine line properties at different line depression levels.\label{ffig2}}
\end{figure}

For all observed or synthetic spectral lines up to 1.1\,$\mu$m, we used the
FTS atlas spectrum \citep{kurucz+etal1984,neckel1999} as a reference. For
spectra at longer wavelength, data from the
BASS2000\footnote{http://bass2000.obspm.fr/solar\_spect.php} spectral
database \citep{delbouille+etal1973} were used. To correct for spectral
intensity gradients in the observations (compare the blue dotted and
black curves in Fig.~\ref{ffig2}), we matched the continuum intensity
of average observed spectra to the corresponding section of the atlas spectra
in wavelength ranges outside (strong) spectral lines. Mismatches outside the
wavelength range relevant for the derivation of the line parameters were
ignored (e.g.~the region at 630.07\,nm in Fig.~\ref{ffig2}). The continuum intensity was normalized as in the case of the reference atlas spectra. The intensity normalization of all synthetic spectra from the simulations was always done fully analogously to the observations, i.e.~with the atlas spectra as reference.
\begin{figure*}
\sidecaption
\resizebox{11.25cm}{!}{\includegraphics{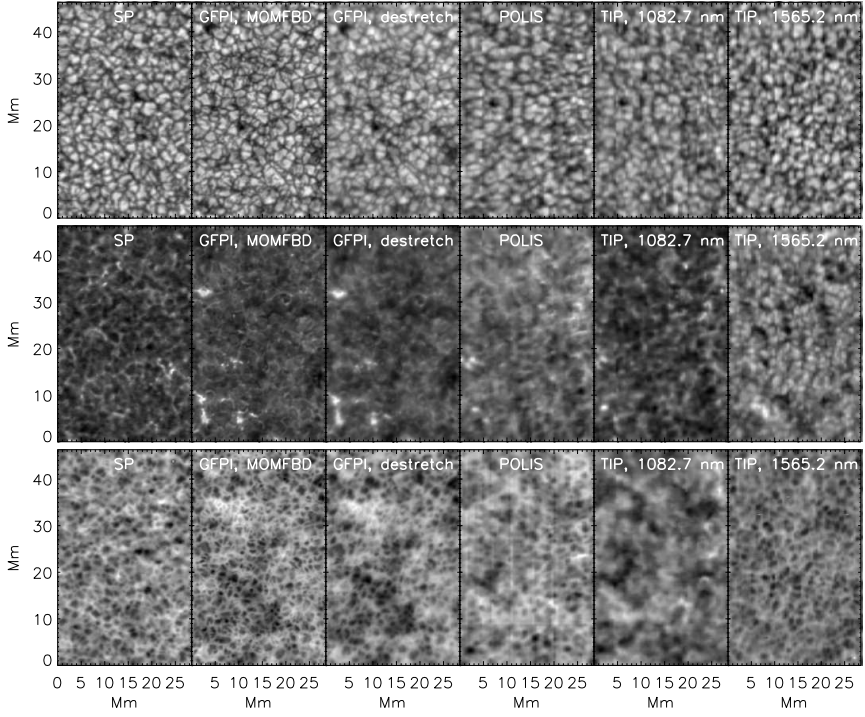}}
\caption{Overview maps representative of the different observational data sets. {\em Top to bottom}:
  continuum intensity $I_{c}$, line-core intensity $I_{\rm core}$ and
  line-core velocity $v_{\rm core}$. {\em Left to right}:  SP, GFPI (MOMFBD),
  GPFI (destretched), POLIS, TIP@1083\,nm, and TIP@1565\,nm. Black/white in the velocity maps denotes motions towards/away from the observer. All images are scaled individually inside their full dynamical range.\label{fig_1}} 
\end{figure*} 
\subsection{Line parameters used}
To compare the thermodynamic properties of the observed and synthetic
spectra, we used the following line parameters: 
\begin{itemize}
\item[1.] The line-core intensity $I_{\rm core}$, i.e.~the intensity at the deepest point of each absorption profile relative to the continuum intensity averaged over the full FOV.
\item[2.] The continuum intensity $I_{\rm c}$ in each profile.
\item[3.] The line depth, or, in other words, the difference $I_{\rm core} - I_{\rm c}$ with both quantities evaluated for each spatially resolved profile.
\item[4.] The line-core velocity $v_{\rm core}$, i.e.~the velocity corresponding to the Doppler displacement of the location of minimal intensity relative to its average location in the FOV.
 \item[5.] The relative line asymmetry, i.e.~the relative difference of the
   areas enclosed between the line profile and the local continuum intensity
   level calculated on the blue or red side, respectively, of the line
   core. Calling those areas $A_b$ and $A_r$, the relative line asymmetry
   is then given by
\begin{equation}
\delta A = \frac{A_b -A_r}{A_b + A_r}\,. \label{eq_1}
\end{equation}
\item[6.] The equivalent width, i.e.~the total area of the absorption
  profile after its normalization to the average value of $I_c$.
\item[7.] The full-width at half-maximum (FWHM) of the line, i.e.~the FWHM of a Gaussian fit to  the intensity profile.
\item[8.] A few line properties at ten equidistant line depression
  levels (as illustrated in Fig.~\ref{ffig2}) between the local continuum and
  the line-core, namely, the line width, the bisector velocity, and the
  intensity value of each of the ten depression levels (called ``bisector
  intensity'' in the following). The intensity levels $I_j (x,y)$
    ($j=1\hdots10$) were calculated by
\begin{eqnarray}
I_j (x,y) =  I_{\rm core} (x,y) + j/10 \cdot \left( I_{\rm c} (x,y) - I_{\rm
    core} (x,y) \right)\,,
\end{eqnarray}
where $x$ and $y$ denote the pixel column and row inside the FOV, respectively. 
This calculation retrieves bisector properties at a constant line
depth, which is assumed to correspond to layers of constant optical depth.
\end{itemize}

Figure \ref{ffig2} shows an example spectrum from the POLIS 630\,nm data, with
some of the relevant ranges and derived quantities. The continuum intensity
was taken as the mean intensity inside the wavelength range limited by the two
dashed lines. For the ground-based 630\,nm data (POLIS, GFPI), we
removed the telluric O$_2$ blend at 630.27\,nm by fitting a Gaussian to it that
was subsequently subtracted. We applied the same routine to observational and
synthetic spectra to compute the corresponding line parameters, with the
wavelength ranges adapted for use with each of the target lines in the
corresponding wavelength regime (557\,nm, 630\,nm, 1083\,nm, 1565\,nm).

Figure \ref{fig_1} shows maps of three of the line parameters ($I_{\rm c}$,
$I_{\rm core}$, $v_{\rm core}$) for the six observational data sets obtained
with the SP, GFPI, POLIS, and TIP. To facilitate a direct
comparison in the figure, we extracted an equal-sized area of about
30$^{\prime\prime}$ by 46$^{\prime\prime}$ from each observation. The two
different approaches to reduce the GPFI data (MOMFBD and destretching, second and third column) have an identical FOV albeit at different
spatial resolutions. The shown sections of the POLIS data at 630\,nm and of
the TIP data at 1082.7\,nm (fourth and fifth column) have been
roughly aligned with one another to pixel precision to again facilitate an
easier visual comparison, i.e.~by obtaining a common FOV. All other data are
neither co-spatial nor co-temporal. The maps show a clear variation in spatial
resolution that approximately decreases from left to right, showing
increasingly smeared structures. The MOMFBD GFPI data have the highest
spatial resolution, the Echelle data at 557\,nm (not shown here, see Fig.~\ref{kernel_2d} later on) have a significantly worse spatial resolution because of the seeing conditions during this observation.

\subsection{Spatial degradation\label{spatdeg}}
To match the spatial resolution of simulations and observations, we determined
suitable convolution kernels to degrade the
simulations to the level of the observations \citep[see,
e.g.][]{danilovic+etal2008,wedemeyer+etal2009,pereira+etal2009,hirzberger+etal2010,afram+etal2011}.
We used a kernel with two components, i.e.~a Gaussian $K_{\rm
  Gauss}(r,\sigma)$ that is assumed to mainly reflect the spatial resolution and a Lorentzian $K_{\rm Lorentz}(r,a)$ that should mimic the stray light
contribution by means of its extended wings \citep[e.g.][]{mattig1983}, where
$r$ denotes the radial distance from a given pixel. $\sigma$ is the variance
of the Gaussian, and $a$ the parameter that regulates the shape of the
Lorentzian. The final convolution kernel $K_{\rm total}(r,\sigma,a)$ is then
given by the convolution of the two kernels: 
\begin{eqnarray}
K_{\rm Gauss}(r,\sigma) = \frac{1}{\sqrt{2 \pi}\sigma} \exp^{-\frac{r^2}{2 \sigma^2}}\\
K_{\rm Lorentz}(r,a) = \frac{1}{\pi}\;\frac{a}{r^2+a^2}\\
K_{\rm total}(r,\sigma,a) = K_{\rm Gauss}(r,\sigma)\ast K_{\rm Lorentz}(r,a)\,,
\end{eqnarray}
where $\ast$ denotes the convolution product.
\begin{table*}
\caption{{\em Top row}: ``best'' degradation kernel parameters $a$ and
  FWHM($\sigma$) for each instrument. {\em Second row}: average spatial stray
  light level $\alpha$ corresponding to each of those ``best'' kernels. {\em
    Bottom rows}:  width of ``best'' kernel at three percentage levels of
  maximum amplitude. \label{tab_stray}}
\begin{tabular}{c|ccccccc}\hline\hline
 & SP & GFPI, MOMFBD & GFPI, destr. & POLIS & TIP@1083\,nm &TIP@1565\,nm & ECHELLE@557\,nm\cr\hline
$a$ / FWHM [$^{\prime\prime}$] & {\bf 0.02 / 0.59} &  {\bf 0.09 / 0.12} & {\bf 0.20 / 0.19}  & {\bf 0.17 / 0.34 } & {\bf 0.04 / 0.96}  & {\bf 0.12 / 0.59}  & {\bf 0.25 / 1.33} \cr
$\alpha$ & {\bf 17\,\%} & {\bf 46\,\%} & {\bf 66\,\%} & {\bf 24\,\%} & {\bf 31\,\%} & {\bf 28\,\%} & {\bf 73\,\%} \cr
 50\,\% &  0.39$^{\prime\prime}$ &     0.34$^{\prime\prime}$ &      0.37$^{\prime\prime}$ &       0.39$^{\prime\prime}$ &       0.42$^{\prime\prime}$ &       0.39$^{\prime\prime}$ &  0.45$^{\prime\prime}$\cr
 25\,\% &  0.58$^{\prime\prime}$ &   0.47$^{\prime\prime}$ &      0.55$^{\prime\prime}$ &      0.58$^{\prime\prime}$ &       0.74$^{\prime\prime}$ &      0.66$^{\prime\prime}$ &  0.87$^{\prime\prime}$\cr
 10\,\% &  0.81$^{\prime\prime}$ &   0.56$^{\prime\prime}$ &       0.81$^{\prime\prime}$ &       0.87$^{\prime\prime}$ &        1.16$^{\prime\prime}$ &     1.00$^{\prime\prime}$ &  1.55$^{\prime\prime}$\cr
\end{tabular}
\end{table*} 

As discussed in, e.g.~\citet[][]{pereira+etal2009}, the parameters $(\sigma,
a)$ of the two kernels cannot be chosen independently. To
prove that and to obtain a (semi-)automatic method for the derivation of the
best parameter values, we quantified the mismatch between observations and
the convolved simulations in the following way. We calculated histograms of the continuum intensity in all pixels in the
FOV of each observation and in the spatially degraded HD simulation
snapshot. Subsequently, we fitted Gaussian functions to all of the histograms
(cf.~Fig.~\ref{int_hist} later on for an example) from either the
observations ($G_{obs}$) or the degraded HD simulation ($G_{degr}$). The
fitted Gaussian curves were always shifted to a fixed reference central value
and normalized by their maximum. The least-square value $\chi^2$ to be
minimized was then defined using two contributions, one related to the
deviation between the two Gaussian fits, and the other related to the
difference between root-mean-squared (rms) continuum intensity contrasts.
For asymmetric distributions the rms value and the width of the fitted
Gaussian can differ significantly because a Gaussian fit mainly reproduces the core of a distribution, whereas the estimator of the rms takes also (far) outliers into account. We used a tenfold increased weighting of the rms contrast to achieve a comparable magnitude for both contributions. The $\chi^2$ was therefore set to
\begin{eqnarray}
\chi^2 = \int_I (G_{\rm degr.}(\sigma,a) - G_{\rm obs.} )^2 dI + 10\cdot | {\rm rms_{\rm degr.} - rms_{\rm obs.} } | \,. \label{eq_chi}
\end{eqnarray}
\begin{figure}
\resizebox{8.8cm}{!}{\includegraphics{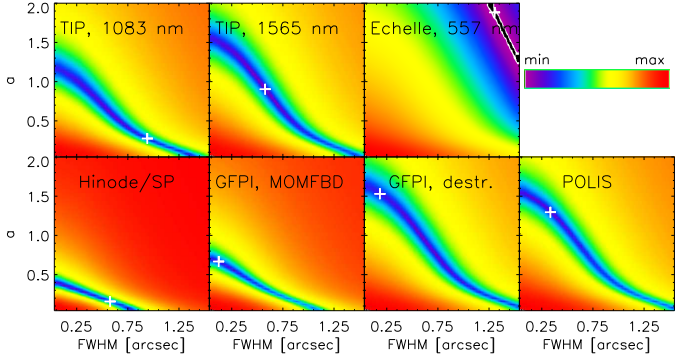}}
\caption{$\chi^2$ values as a function of FWHM ($= 2\,\sqrt{2\,\log 2}\,\sigma$) and $a$. {\em Bottom row, left to right}:  SP, GFPI (MOMFBD), GFPI
  (destretched) and POLIS. {\em Top row}: TIP@1083\,nm, TIP@1565\,nm and
  Echelle@557\,nm. The white crosses indicate the parameters
  corresponding to the ``best'' kernels. \label{chifig}} 
\end{figure}

The $\chi^2$ was calculated for values of the FWHM of the $K_{\rm Gauss}$ kernel (Eq.~2) between 0\farcs12 and 1.5$^{\prime\prime}$ and for $a$ in Eq.~(3) between 0 and 2, using for $G_{obs.}$ and rms$_{\rm obs.}$ the different observational data sets separately. The determination of the minimum of the resulting $\chi^2$-surface turned out to be a degenerate problem: as indicated by the blue/black colored area in
Fig.~\ref{chifig}, for (nearly) each value of $\sigma$ a suitable value of $a$ can be found that yields a similarly small $\chi^2$. However, it is the parameter $\sigma$ that has the strongest influence on the resulting spatial resolution. We therefore selected ``best'' kernels by choosing the $\sigma$ value that provided the closest match in spatial resolution when visually comparing the observations and the degraded HD simulation (cf.~Fig.~\ref{kernel_2d} later
on). Having fixed $\sigma$, the minimum of the $\chi^2$ function provides the corresponding $a$. The white crosses in Fig.~\ref{chifig} denote the final choice of $(\sigma,a)$ for each observation (listed in the top row of Table
  \ref{tab_stray}).

\begin{figure}
\resizebox{8.8cm}{!}{\includegraphics{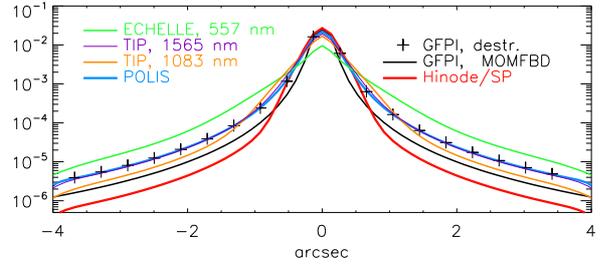}}
\caption{``Best'' degradation kernels $K_{\rm total}$ for GFPI MOMFBD (black solid line), GFPI destretched (black crosses), Hinode/SP
  (thick red line), POLIS (thick blue line), TIP@1083\,nm (orange line), TIP@1565\,nm (purple line) and Echelle@557\,nm (thick green line). \label{kernel_fig}}
\end{figure}

Figure \ref{kernel_fig} shows the final ``best'' kernels for matching the convolved HD simulation with the various observations. The shape of the kernels fits the expectations when considering the visual impression of the data properties (Fig.~\ref{fig_1}). Namely, the convolution kernels for the ground-based spectrograph data and the destretched GFPI spectra have broader cores and higher wings than the SP and MOMFBD GFPI data, which leads to the lower contrast and worse spatial resolution of the former. The GPFI MOMFBD kernel has the smallest core of all, in agreement with its intensity map in Fig.~\ref{fig_1} that shows more small-scale fine-structure than any of our other observations.

The convolution kernels can be used to estimate the generic stray light level
in the observations. We calculated the total spurious intensity introduced
into the central pixel of the simulation box from its surroundings
\citep[cf.][BE11, their Eq.~8]{beck+etal2011a} as
\begin{equation}
I_{\rm stray}(x,y, \lambda) = \sum_{x^\prime\neq x,y^\prime\neq y}
K(x-x^\prime,y-y^\prime) \cdot I(x^\prime,y^\prime, \lambda)\,, 
\end{equation}
{\bf where $I(x^\prime,y^\prime, \lambda)$ are the synthetic spectra of the HD simulation at full resolution (HD-FR) after re-sampling them and the kernel to the spatial sampling of the respective observation, while $x$ and $y$ denote the pixel column and row inside the re-sampled FOV, respectively.}
simulation at full resolution (HD-FR) and $x$ and $y$ denote the pixel
  column and row inside the FOV, respectively.

The generic spatial stray light level $\alpha$ is then given by the
value of $I_{\rm stray}(x,y, \lambda)$ in a continuum wavelength range. The
second row of Table \ref{tab_stray} lists the average stray light level
for the various instruments expected for QS conditions. {\bf All spatially under-sampled spectrograph data (SP, POLIS, TIP) yield similar values of about 20--30\,\%, with little to no dependence on the spatial resolution.} {\bf A clear reduction of stray light by 20\,\% is seen for the MOMFBD GFPI spectra relative to the destretched GFPI data, because in the deconvolution process some inverse kernel was already applied.} Our generic stray light estimate encompasses
  both the effects of stray light from scattering on optics and from
  resolution effects caused by, e.g.~the finite aperture size of the
  corresponding telescope \citep[cf.~for instance][]{danilovic+etal2008}. The
  contribution of resolution effects to stray light is present even for
  diffraction-limited observations and dominates the shape of the core of the
  kernel. For comparison, \citet[][]{wedemeyer+etal2009} determined an
  effective value of $\alpha$ of 33\,\% for the blue channel of the SOT/BFI
  onboard Hinode, of which about 8\,\% should be caused by stray light and
  the rest by resolution effects \citep[cf.][]{wedemeyer2008}. For the SP,
  \citet{danilovic+etal2008} required a stray light level of about 5\,\% in
  addition to the theoretical point spread function (PSF) from the resolution effects to match the
  continuum contrast of their simulations and SP observations. Stray light
  estimates for POLIS are listed in BE11, with a lower limit of 10\,\% in
  umbrae of sunspots and about 24\,\% in QS. 

The width of the kernels at three percentage levels of the maximum kernel amplitude is listed in the lower rows of Table \ref{tab_stray}. The FWHM of the convolution kernel (row labelled 50\,\%) does not seem to be a good proxy for the spatial resolution because its variation across the different data sets is clearly too small, with a similar value of about 0\farcs4 for all instruments. Only at the lower two levels (25\,\%, 10\,\%) does the difference in quality become prominent, with observations of worse spatial resolution showing significantly broader kernels. For the ground-based slit-spectrograph data (TIP, POLIS, Echelle; cf.~the fifth row of Table \ref{table1} for the sampling-limited resolution), we accordingly used the width at 25\,\% of the maximum kernel amplitude as an
estimate of the spatial resolution.
\begin{figure}
\resizebox{8.8cm}{!}{\includegraphics{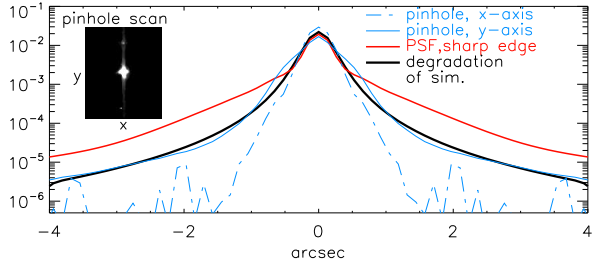}}
\caption{Estimates of the PSF for POLIS from the degradation of the HD
  simulation (thick black line), a measurement of a knife-edge function (red
    line) and a scan across the pinhole in F1 (blue/blue-dot-dashed lines: cut along y/x). The inset at the upper left shows the intensity map of the pinhole scan.\label{kernel_polis_fig}}
\end{figure}

The derivation of the degradation kernel by a convolution of simulations is an
indirect method that implicitly assumes that the simulation can be used as
reference for the true spatial intensity distribution. In the case of
ground-based observations, the actual degradation has three contributions: the
degradation by seeing fluctuations, by the telescope optics, and by the
(instrument) optics behind the focal plane of the telescope (F1)
\citep[e.g.][]{mattig1971,mattig1983}. The latter part, the instrumental
PSF, can also be directly measured by some methods,
usually using some type of (artificial) occulting disc in the telescope focal
plane
\citep[cf.][]{briand+etal2006,wedemeyer2008,beck+etal2011a,loefdahl+scharmer2012}.

In the case of POLIS, we have an estimate of the PSF from an observation with
a partly blocked FOV in F1 (BE11) and from a scan of a pinhole in F1 taken in
April 2011. Figure \ref{kernel_polis_fig} compares the convolution kernel
derived for the HD simulation with the direct measurements of the PSF. The
degradation of the HD simulation and the pinhole scan along the slit ($y$
axis) yield nearly identical kernels. The small oscillation around the simulations' PSF likely represents the Airy pattern. The knife-edge measurement used in BE11 yields the same core, but has quite elevated wings compared to the other two methods. The cut in the scanning direction ($x$) yields a smaller kernel that drops to zero faster than the one for the cut in $y$, because only light that is scattered already in front of the spectrograph is detected. The observation of a
pinhole seems to be a practical approach for an observational
determination of the instrumental PSF
\citep[cf.][]{loefdahl+scharmer2012,puschmann+etal2012c}, while the good
match of the measured PSF with the one derived from degrading the
simulation\footnote{For the SP, the kernel derived from the degradation
  of the HD simulation matches a theoretical calculation of the PSF of the
  SOT (A.Asensio, private communication).} suggests the latter as method of choice to obtain a PSF
estimate otherwise.

Figure \ref{int_hist} shows the quality in the reproduction of the observed
intensities by the degradation of the simulations with the respective ``best''
kernels. The intensity histograms of the data with high spatial resolution are asymmetric, with an extended tail towards intensities above unity (SP and GFPI MOMFBD data in the top row of Fig.~\ref{int_hist}). The 
deviation of the average value from unity seen in some cases is caused by the
intensity normalization of the spectra to the FTS atlas, because some of the
chosen ``continuum'' wavelength ranges happened to fall in the far wings of
strong lines (see Fig.~\ref{ffig2}). The rms contrasts were matched to $\sim
0.3$\,\% (cf.~Table \ref{big_table} later on) and the intensity histograms of the observations are matched adequately.
\begin{figure}
\resizebox{8.8cm}{!}{\includegraphics{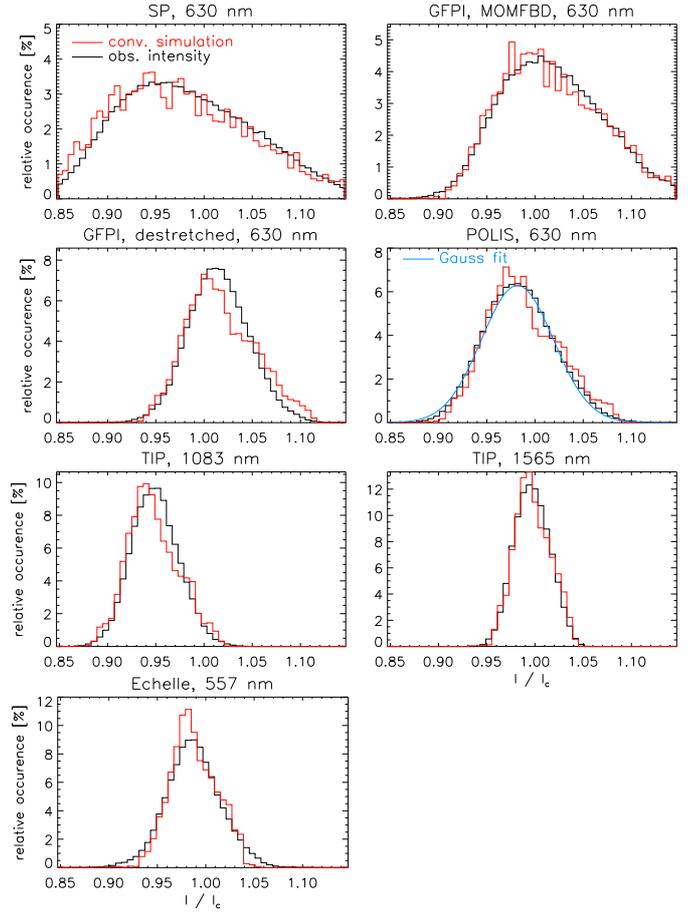}}
\caption{Intensity histograms for the observations (black lines), and (red lines) for the HD simulation spatially convolved with the corresponding ``best''
  kernel. Instrument and wavelength are given at the top of each panel. The blue line in the panel corresponding to the POLIS data
  denotes the Gaussian fit to the observed intensity distribution.\label{int_hist}}
\end{figure}
\begin{table*}
\caption{Parameters of the spectral degradation from FTS atlas to observations.\label{tab_spec}}
\begin{tabular}{c|cc|cc|cc|c|cc|ccccc}\hline\hline
instr. & \multicolumn{2}{c|}{SP} & GFPI,  MOMFBD & GFPI, destr. &
\multicolumn{2}{c|}{POLIS}  &  \multicolumn{3}{c|}{TIP}  & \multicolumn{3}{c}{Echelle} \cr
line  & 630.15 &  630.25 &  630.25 &  630.25& 630.15 &  630.25 & 1082.7 &
1564.8 & 1565.2 &  557.6 & 557.7 & 557.8 \cr\hline 
$\beta$ [\%]& 2.9 & 2.4 & 9.7 & 10.3 & 20.3 & 20.5 & 8.1  & 9.3 & 19.5 & 1.6 & 2.6 & 3.4 \cr
$\sigma^\prime$ [pm] & 1.54 & 1.95 & 2.08 & 2.01 & 2.62 & 2.89 & 7.05 & 5.97 & 4.36 & 2.08 & 1.98 & 1.78 \cr
$\sigma^\prime$ [km/s] &  0.73&     0.93&     0.99&     0.96&      1.25&      1.38&      1.95&      1.14&     0.84 & 1.12 & 1.07 & 0.96 \cr
\end{tabular}
\end{table*}
\begin{figure}
\resizebox{8.8cm}{!}{\includegraphics{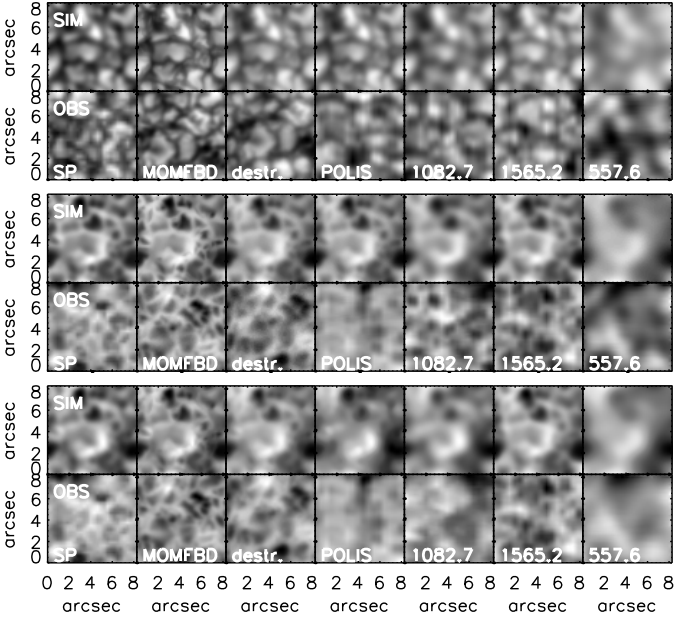}}
\caption{Comparison of observations ({\em lower row} in each panel) with the
HD simulation ({\em upper row} in each panel) convolved using each
observation's ``best'' kernel. {\em Top to bottom}: $I_c$, $v(50\,\%{\rm\,
  line\,depression})$ and $v_{\rm core}$. {\em  Left to right}: SP, GFPI(MOMFBD), GFPI(destretched), POLIS, TIP@1083\,nm, TIP@1565\,nm and Echelle@557\,nm.\label{kernel_2d}}
\end{figure}

The general match between observations and simulation
throughout a 2D FOV is shown in Fig.~\ref{kernel_2d}. We cut out a 6
Mm\,$\times$\,6 Mm square from each observation for a visual comparison to
the HD simulation convolved with the 
corresponding best kernel. The spatial patterns and the spatial resolution in
$I_c$ (top panel of Fig.~\ref{kernel_2d}), the velocity at 50\,\%
line depression (middle panel), and the line-core velocity (bottom
  panel) are well reproduced for all observations. Note, however, that the
simulation data used for this comparison were already spectrally
degraded as well (see the next section). Simulation data that were only
spatially degraded showed some artifacts in the line-core velocity for all
spectral lines except the low-forming \ion{Fe}{i} lines at 1565\,nm. In case
of the GFPI MOMFBD data (second column in Fig.~\ref{kernel_2d}) it would
be impossible to distinguish the observed and synthetic maps if the
corresponding labels were absent. The spatial resolution and the rms contrast
of the shown section of the 557\,nm data is higher than throughout the rest of the observed FOV at this wavelength because the part presented in the figure belongs to the region around the AO lock point.
\subsection{Spectral degradation\label{spec_deg}}
The spectral degradation was determined by matching FTS
spectra and average observed profiles by adding a wavelength-independent
stray light offset $\beta$ to the atlas data and subsequently convolving them with a Gaussian $G$ of width $\sigma^\prime$ \citep{allendeprieto+etal2004,cabrerasolana+etal2007} according to
\begin{equation}
I_{\rm FTS, degraded} (\lambda) = \frac{I_{\rm FTS}  (\lambda) +
  \beta}{1+\beta}\ast G(\sigma^\prime, \lambda)\,.
\end{equation}
The value of $\beta$ only refers to spectral stray light produced by scattering inside the spectrometer, whereas the spatial stray light produced in front of the slit or the instrument entrance does not contribute to the spectral degradation (BE11). The squared deviation between the degraded FTS profile and the observed average profile as a function of $\beta$ and
$\sigma^\prime$ yields again a matrix whose minimum indicates the best match
of the profiles \citep[e.g.][their Fig.~A.1]{cabrerasolana+etal2007}. We
applied the procedure to all strong observed spectral lines in the respective
wavelength regimes, in addition to those used for deriving the line
parameters. The corresponding best-match parameters $\beta$ and
$\sigma^\prime$ are listed in Table \ref{tab_spec}. For the spectral
degradation of the simulations in those cases, where multiple lines were
observed with the same instrument and within the same spectral region (e.g.~1564.8\,nm and 1565.2\,nm), we used the {\it average} values of $\beta$ and
$\sigma^\prime$. The spectral stray light levels range between $\sim$ 2\,\% (Echelle, SP) and 10\,\% or more (TIP, POLIS). The velocity equivalents of the spectral broadening are about 1\,kms$^{-1}$ in most cases.

After applying the spatial degradation, or spatial and spectral degradation to the synthetic spectra of the HD simulation (denoted in the following by HD-SPAT and HD-SPAT-SPEC, respectively), we determined the line parameters, and
inverted the spectra with the settings described in the next section. The
noise level in the observed intensity spectra is of the order of $10^{-3}$ of $I_c$ and should have a negligible impact on the analysis results as long as only
Stokes $I$ is involved. We therefore did not add noise to the synthetic
spectra.
\subsection{Inversion setup}
Using the SIR code \citep[Stokes inversion by response functions;][]{cobo+toroiniesta1992}, we inverted the spectra of
all observations (SP, GFPI, POLIS, TIP), as well as the synthetic 630\,nm
spectra in the HD-FR, HD-SPAT, and HD-SPAT-SPEC. The spectral lines in our
set of observations qualify as medium to strong lines and therefore provide
information over an extended height range from the continuum forming layers
up to $\log\tau \approx -1.5$ to $-2$
\citep[cf.][]{allendeprieto+etal2001b,cabrera+bellot+iniesta2005}. The
inversion was done with a fixed generic spatial stray light level $\alpha$ of 20\,\% for all the observations, and for the HD-SPAT and HD-SPAT-SPEC cases, whereas $\alpha$ was set to zero for the inversion of the synthetic profiles at full
resolution. Only one component (filling factor 100\,\%) was used in
the inversion that was set
to be either field-free or magnetic depending on whether the polarization
degree in the spectra exceeded a specific threshold. This setup forced the
inversion code to use just the temperature stratification to reproduce the
observed Stokes $I$ profile on each pixel, and hence the temperature in the
horizontal plane also follows the spatial intensity variation.
\begin{figure*}
\resizebox{17.6cm}{!}{\includegraphics{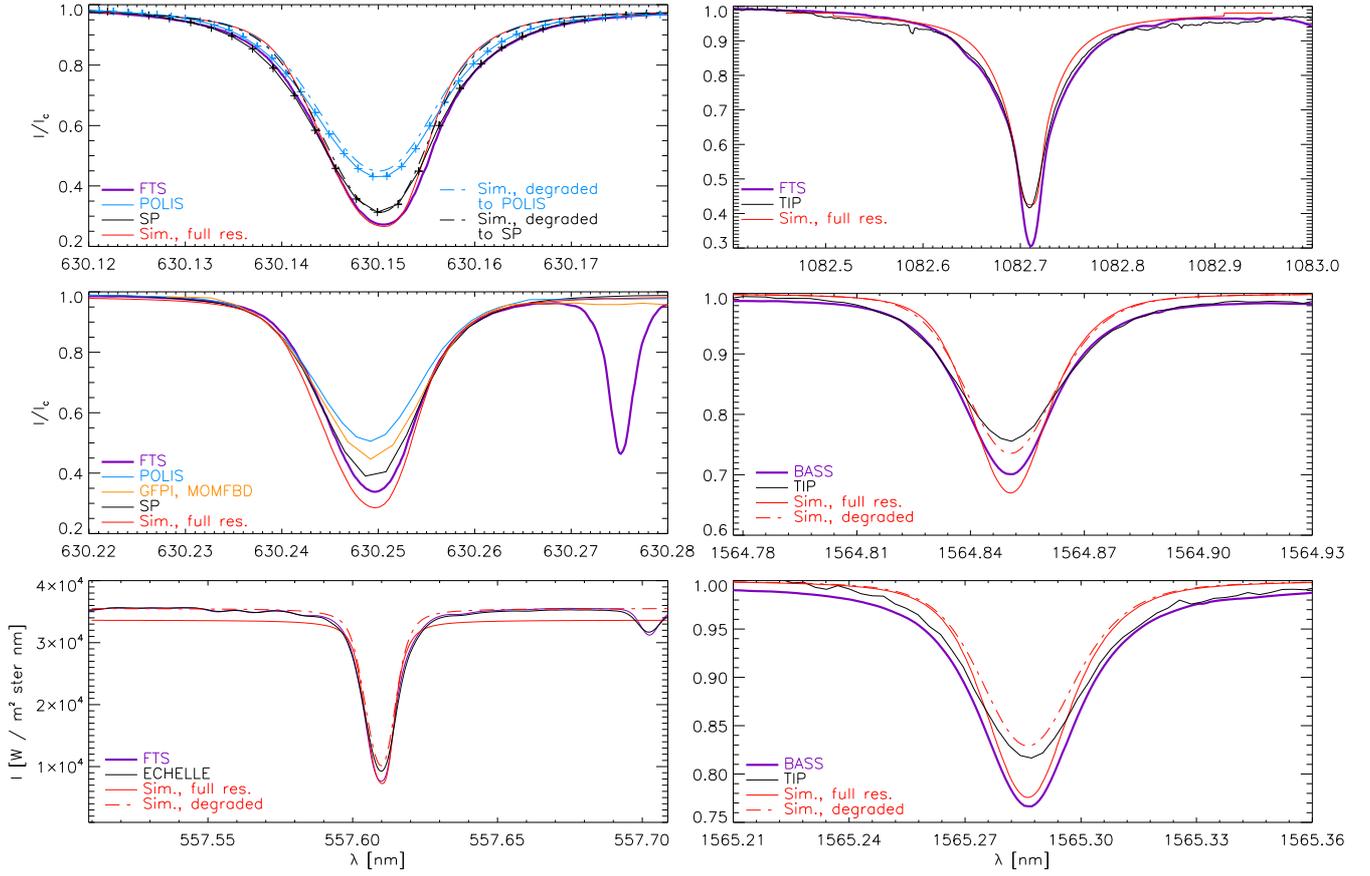}}
\caption{Spatially averaged profiles of the 630\,nm ({\em left top two rows}),
  the 557\,nm range ({\em bottom left}), and the near-IR wavelength ranges
  ({\em right column}). For the 630.15\,nm line, the spectral sampling of the SP and POLIS is indicated by crosses. The thick purple curves show the spectra in the FTS or BASS atlas, depending on the wavelength range of interest. For 557.6\,nm, 630.15\,nm, 1564.8\,nm and 1565.2\,nm the profiles from the HD-SPAT-SPEC are over-plotted with dash-dotted lines. In the plot at 557.6\,nm, absolute flux units are used and the normalization of the HD-FR spectra to the FTS was undone.\label{av_prof}}
\end{figure*}
\begin{figure*}
\resizebox{17.6cm}{!}{\includegraphics{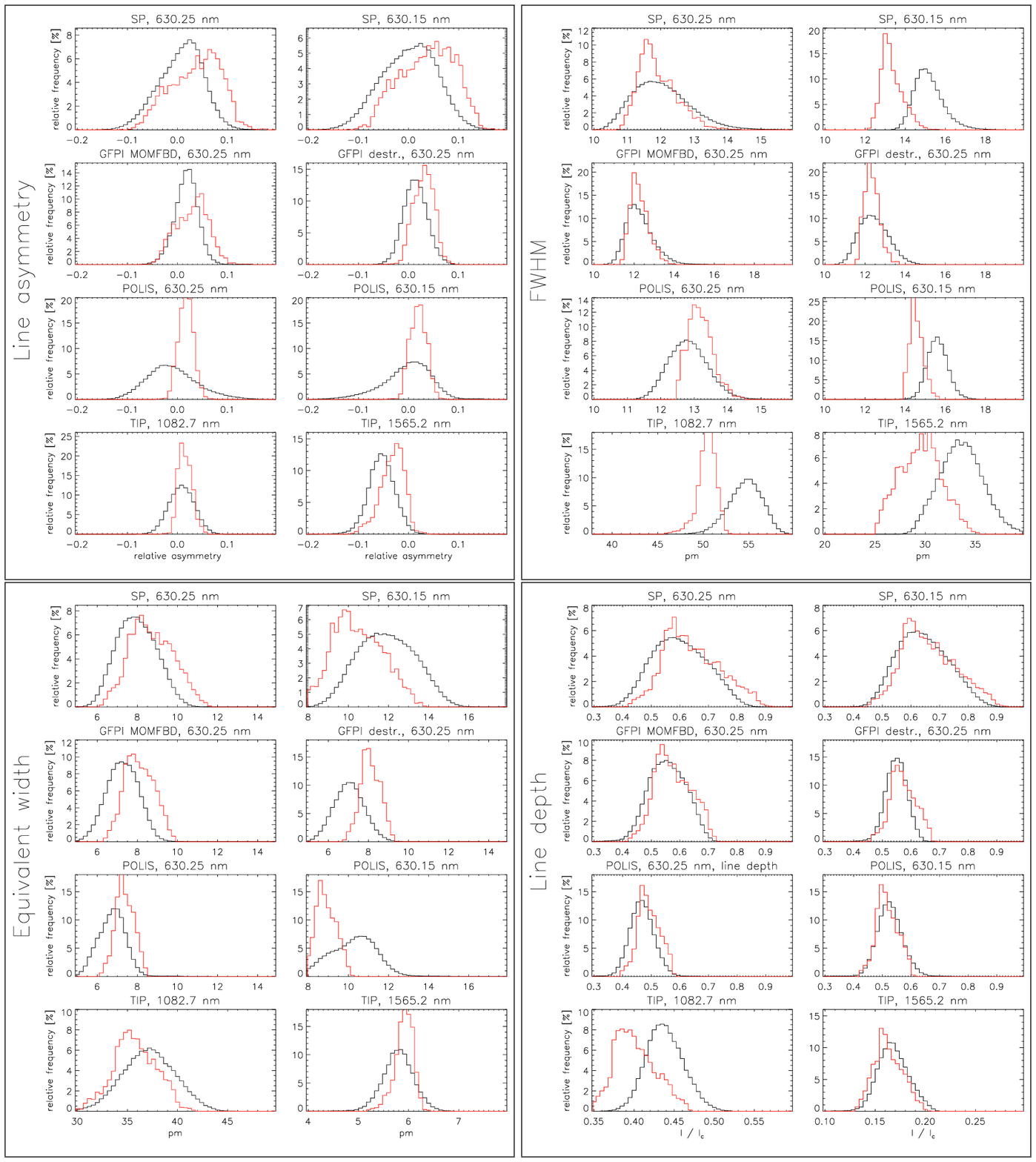}}
\caption{Histograms of line parameters derived for HD-SPAT-SPEC and observations. {\em Top left panel}: line asymmetry. {\em Top right panel}: FWHM.  {\em Bottom left panel}: equivalent width. {\em Bottom right panel}: line depth. The black line denotes the histogram of the relevant parameter for a given observational data set, the red line the one for the degraded simulation. Instrument and wavelength are denoted at the top of each panel.\label{fig_stat_rest}}
\end{figure*}

Inversion setups with a variable stray light contribution or multiple
inversion components do not reproduce the spatial variation of temperature
well because of the interplay between temperature, stray light contribution,
and/or filling factors \citep[e.g.][]{martinez+etal2006}. Effective magnetic
filling factors are about 10\,\% for ground-based observations at
1$^{\prime\prime}$ resolution
\citep[e.g.][]{martinez+etal2008,beck+rezaei2009} and reach 20\,\% for SP
data \citep{orozco+etal2007a}. When using a variable ``stray light
contribution'' in a one-component inversion as the only factor taking the
unresolved structure of the magnetic field into account, inversion codes will
commonly tend to set it to unreasonably large (cf.~Sect.~\ref{spatdeg})
values of up to 80\,\%, because under such an assumption, this factor has to
reproduce both the real stray light and the magnetic filling factor inside a
pixel \citep[but see also][]{lagg+etal2010,marian+etal2012}. Our choice of the inversion
setup yields an acceptable fit to the Stokes $I$ profiles (indicating a
reasonable choice of $\alpha$) while the polarization signal in Stokes $QUV$
is less well reproduced because of our one-component assumption. This has no
impact on the current study because we do not make use of the retrieved
information on the magnetic field parameters.

We chose the Harvard Smithsonian Reference Atmosphere
\citep[HSRA,][]{gingerich+etal1971} as initial temperature model. We used
only two nodes ($\equiv$\,variation of the initial model by linear gradients)
for the temperature stratification. From the inversion, we retrieved $v_{\rm
  mac}$, $v_{\rm mic}$, and the temperature stratification $T$ as characteristic
thermodynamic parameters.
\section{Results\label{sec_res}}
\subsection{Average line profiles of HD simulation and observations}
Figure \ref{av_prof} shows the spatially averaged profiles from the
observations and from the HD simulation, as well as the atlas profiles. The average spectra from the HD-FR are expected to reflect the averaged thermodynamic conditions in the QS (even if with some limitations), and thus provide a good fit to the  observed FTS spectrum. Possible sources of deviation are the
lack of magnetic broadening, a somewhat inadequate choice of thermodynamic
parameters in the simulation, and the transition parameters and solar abundances used in the spectral synthesis. 

The comparison of the FTS profile (thick purple curve in
Fig.~\ref{av_prof}) with the HD-FR spectra (red line) shows a good
match for the 630.15\,nm line, an acceptable match for 630.25\,nm, and from moderate to large deviations for the rest of the
spectral lines, mainly because at the corresponding wavelengths (e.g.~557.6\,nm, 1082.7\,nm, 1565\,nm), the HD-FR yields line profiles with
narrower wings than present in the FTS spectrum. This is likely due to the collisional broadening approximation employed in the spectral synthesis. For abundance determinations, spectral synthesis codes such as NICOLE \citep{socasnavarro+etal2000} can be employed that allow for the use of the more reliable collisional treatment developed by Anstee, Barklem, and O'Mara \citep[ABO, see][]{barklem+etal2000} as well as for the inclusion of non-LTE effects. The small deficiency ($\lessapprox 5$\,\%)  in continuum intensity caused by the slightly too low effective temperature of the simulations (cf.~Sect.~\ref{sec_sim}) is demonstrated in the bottom left panel of Fig.~\ref{av_prof}. The normalization of all spectra to the atlas reference spectra removes it efficiently.

The spectral degradation determined in the previous section transforms the
FTS spectra to the average profile of each observation. It therefore produces
a good match of simulations and observations only when the spectra from the
simulations at full resolution resemble the FTS to start with. This is
especially the case for the 630.15\,nm line, but also for the rest of the
lines the match between the degraded HD simulation and the observations is
improved by the spectral degradation.
\begin{figure}
\resizebox{8.8cm}{!}{\includegraphics{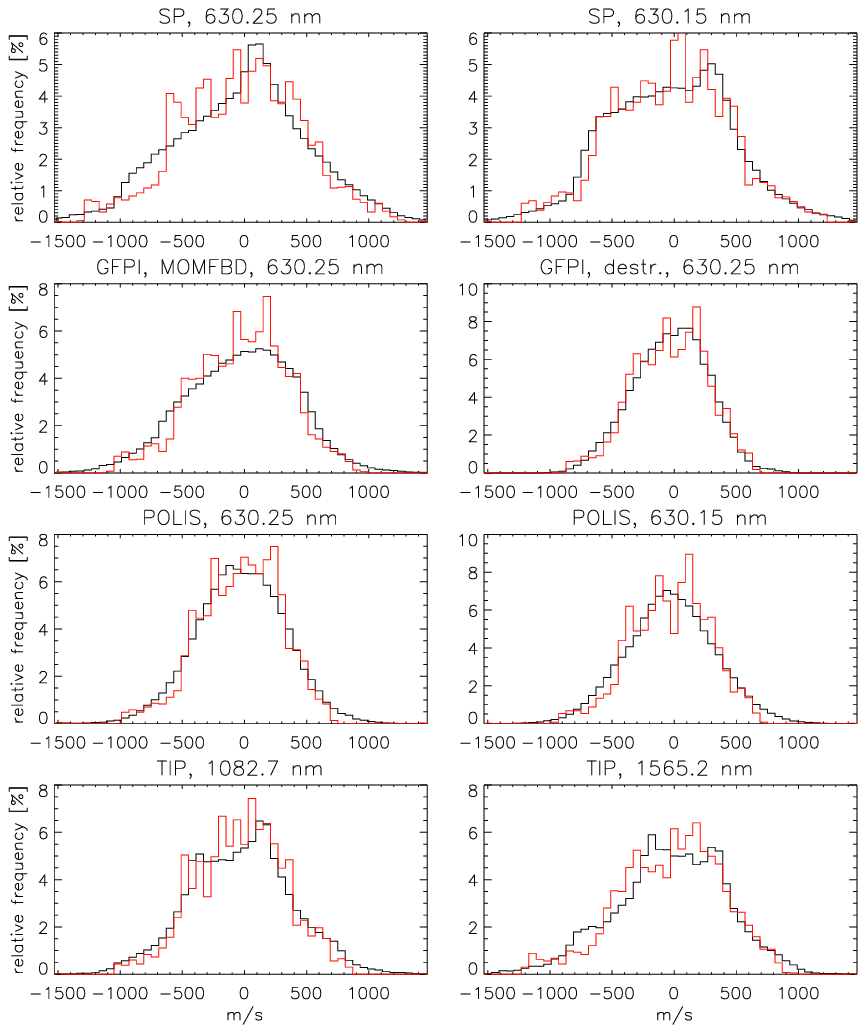}}
\caption{Histograms of line-core velocity in HD-SPAT-SPEC and
    observations. The black line denotes the histogram of the
  parameter in the observation and the red line that in the HD-SPAT-SPEC. Instrument and central wavelength of relevance are
  specified in the title of each sub-panel. \label{fig_stat_velo}}  
\end{figure}
\begin{table*}
\caption{Line parameters in observations and simulations.\label{big_table}}
\begin{tabular}{ccccccccc}\hline\hline
type & line & rms $I_c$ & rms $v(50\%)$& rms $v_{\rm core}$ & line depth & core intensity & asymmetry &
eq.~width\cr
units & nm & \%  & ms$^{-1}$ & ms$^{-1}$ & $<I_c>$ & $<I_c>$ & $\delta A$ & pm\cr\hline
\multicolumn{8}{c}{ECHELLE}\cr\hline
HD FULL RES&557.6 &  18.1 & 890 &  968 &  0.830 $\pm$ 0.225 &  0.203 $\pm$ 0.032 & -0.033 $\pm$  0.229 & 9.71 $\pm$ 2.80 \cr
HD SPAT & 557.6 & 2.4 & 247  & 173 &  0.764 $\pm$ 0.035& 0.215 $\pm$ 0.009  & 0.099 $\pm$ 0.031 &   9.65 $\pm$      0.35\cr
HD SPAT+SPEC &557.6 & 2.4 &  234 &  227 & 0.689 $\pm$  0.030 & 0.281 $\pm$ 0.012 & 0.036 $\pm$  0.012 &   9.39 $\pm$  0.34\cr
OBS\tablefootmark{1}& 557.6 & 2.6 & 282 &       316 & 0.679 $\pm$ 0.033 & 0.257 $\pm$ 0.016 &  0.020 $\pm$  0.027   &    12.27 $\pm$      0.43\cr\hline
\multicolumn{8}{c}{POLIS}\cr\hline
HD FULL RES & 630.15 & 15.2 & 1102 & 986 &   0.760   $\pm$   0.190 &      0.254  $\pm$   0.034 & -0.032 $\pm$   0.211 &       10.76   $\pm$   2.72 \cr
HD SPAT & 630.15 & 3.7 &      362 & 224 &     0.704 $\pm$  0.052 & 0.266 $\pm$    0.012 & 0.091 $\pm$ 0.052 &  10.72 $\pm$     0.63 \cr
HD SPAT+SPEC &630.15 & 3.7  &  288 &  315 &      0.523 $\pm$ 0.037  &    0.450 $\pm$     0.013&   0.023 $\pm$  0.016  &  8.95 $\pm$      0.53\cr
OBS & 630.15 & 3.7 &  392 &    358 &      0.530  $\pm$   0.059 &      0.425  $\pm$   0.034 & -0.006   $\pm$  0.048 &       10.23   $\pm$    1.57 \cr\hline
HD FULL RES & 630.25 & 15 &1097  & 821 &      0.756  $\pm$    0.201 &     0.269 $\pm$    0.049 & -0.009 $\pm$     0.225   &  8.98   $\pm$    2.35 \cr
HD SPAT & 630.25 & 3.7& 365 &  231&     0.688 $\pm$    0.056&     0.285 $\pm$
0.018& 0.084 $\pm$    0.050& 8.94 $\pm$      0.54\cr
HD SPAT+SPEC & 630.25 & 3.7 &  474 & 322 &      0.498 $\pm$  0.038 & 0.481 $\pm$ 0.015 &0.021 $\pm$ 0.015 & 7.46 $\pm$ 0.45\cr
OBS & 630.25 & 3.7 & 381 & 353   &        0.472  $\pm$  0.056 &    0.503   $\pm$
0.038 & -0.007  $\pm$  0.050   &   6.76   $\pm$   0.93 \cr\hline
\multicolumn{8}{c}{SP}\cr\hline
HD SPAT & 630.15 & 7.2& 594 & 360&     0.717 $\pm$    0.098&     0.262 $\pm$
0.019&    0.087 $\pm$     0.103& 10.74 $\pm$      1.24 \cr
HD SPAT+SPEC & 630.15  & 7.2 & 455 &  464 &   0.670 $\pm$     0.093 & 0.307 $\pm$ 0.023 &  0.041 $\pm$ 0.053 &  10.57 $\pm$       1.24\cr
OBS & 630.15 & 6.9 &  619 & 498 &     0.650  $\pm$   0.087  &    0.307 $\pm$   0.034 & 0.004 $\pm$   0.055   &   11.99  $\pm$    1.40 \cr\hline
HD SPAT &  630.25 & 7.2& 591 & 373&     0.703 $\pm$     0.105&     0.280 $\pm$    0.028& 0.080 $\pm$    0.096&  8.96 $\pm$      1.07 \cr
HD SPAT+SPEC & 630.25  & 7.2 &   607 &  483 &  0.649 $\pm$  0.097 &  0.333 $\pm$ 0.030 &     0.040 $\pm$ 0.049 &  8.81 $\pm$       1.06\cr
OBS & 630.25 & 6.9 &  614 & 537 &   0.606  $\pm$   0.094  &    0.374  $\pm$   0.040 & 0.011 $\pm$  0.044  &     8.06  $\pm$     0.97 \cr\hline
\multicolumn{8}{c}{GFPI  MOMFBD}\cr\hline
HD SPAT & 630.25& 5.3 & 474 &  271&     0.693 $\pm$    0.076&     0.283 $\pm$
0.022& 0.081 $\pm$    0.074&   8.94 $\pm$  0.78 \cr
HD SPAT+SPEC & 630.25 & 5.3 &  511&  379& 0.583 $\pm$ 0.063 & 0.396 $\pm$ 0.022 & 0.033 $\pm$ 0.032 &  8.20 $\pm$ 0.72 \cr
OBS & 630.25 & 5.3& 443 & 439 &      0.564 $\pm$     0.064 &      0.460   $\pm$ 0.031 & 0.022   $\pm$   0.023 &      7.36   $\pm$    0.79 \cr\hline
\multicolumn{8}{c}{GFPI destretched}\cr\hline
HD SPAT & 630.25  & 3.7 &      356& 224 &     0.701 $\pm$    0.055 & 0.291 $\pm$    0.018   & 0.083 $\pm$ 0.048 &  9.11 $\pm$     0.54 \cr
HD SPAT+SPEC & 630.25  & 3.7 &  474 &      298 & 0.579 $\pm$ 0.044 & 0.399 $\pm$  0.017 & 0.034 $\pm$ 0.020 & 8.19 $\pm$ 0.48\cr
OBS & 630.25  & 3.2 &      283& 300 &     0.556 $\pm$ 0.037 & 0.464 $\pm$    0.022 &    0.018 $\pm$ 0.023 &      7.13 $\pm$     0.74 \cr \hline
\multicolumn{8}{c}{TIP}\cr\hline
HD FULL RES &1082.7 & 9.0 & 1246 & 1539  &   0.484  $\pm$    0.084 &      0.427  $\pm$
0.016 & -0.024    $\pm$   0.125    &   39.76    $\pm$    8.38 \cr
HD SPAT & 1082.7 & 2.7 &      413 &  517 &     0.475 $\pm$    0.030 &     0.433 $\pm$   0.003 &    0.076 $\pm$ 0.061 &      39.67 $\pm$      2.35\cr
HD SPAT+SPEC &1082.7 & 2.7 &  405 &  357 &  0.403 $\pm$  0.026 &  0.493 $\pm$ 0.006 &  0.019 $\pm$ 0.013 &  35.93 $\pm$       2.13\cr
OBS &1082.7 & 2.5 & 398 & 415 & 0.463 $\pm$ 0.024 & 0.438 $\pm$ 0.014 & 0.045 $\pm$ 0.025 & 38.03 $\pm$ 2.66\cr\hline
HD FULL RES &1564.8 & 7.9& 1347 & 1118 &    0.397   $\pm$  0.093 &      0.603  $\pm$
0.042 & 0.001   $\pm$  0.167 &       9.57  $\pm$    1.58 \cr
HD SPAT &  1564.8 & 1.9 &      554&  378 &     0.318 $\pm$    0.031 &     0.666 $\pm$    0.022 &   -0.016 $\pm$ 0.030 &      9.53 $\pm$     0.32\cr
HD SPAT+SPEC &1564.8 & 1.9 & 347&      385 &     0.259 $\pm$ 0.021 &  0.731 $\pm$ 0.015 & -0.025 $\pm$  0.020 &  8.71 $\pm$      0.29 \cr
OBS &1564.8 &  1.8 & 397 & 431 & 0.227 $\pm$ 0.021 & 0.760 $\pm$ 0.017 &  -0.033
$\pm$ 0.027 & 8.87 $\pm$     0.36 \cr\hline
HD FULL RES &1565.2 & 7.9 & 1453& 1292 &  0.273 $\pm$   0.060 & 0.718 $\pm$ 0.041& 0.005 $\pm$ 0.137& 7.15 $\pm$     1.06 \cr
HD SPAT &  1565.2 &   1.9 &      563&  426&  0.213 $\pm$    0.021 &     0.772 $\pm$    0.018 &   -0.027 $\pm$ 0.032 &      7.13 $\pm$     0.22 \cr
HD SPAT+SPEC &1565.2 & 1.9 & 367 & 414 &  0.165 $\pm$ 0.014  & 0.826 $\pm$ 0.013 & -0.029 $\pm$  0.024 &  5.94 $\pm$      0.18 \cr
OBS &1565.2 &  1.8 & 405 & 445 & 0.170 $\pm$ 0.015 & 0.822 $\pm$   0.016 & -0.050
$\pm$ 0.027 & 5.83 $\pm$     0.31 \cr\hline
\multicolumn{8}{c}{Simulation, MHD\tablefootmark{2}, full resolution}\cr\hline
MHD FULL RES& 630.25, 50 G & 14.0 & 1067 & 856&     0.740 $\pm$     0.192&     0.283 $\pm$
0.088& -0.023$\pm$     0.214& 8.95 $\pm$      2.19 \cr
MHD FULL RES&630.25, 100 G & 14.1 & 1058 & 909&     0.716 $\pm$     0.207&     0.305 $\pm$
0.116&   -0.023 $\pm$     0.214& 8.80 $\pm$      2.23\cr
MHD FULL RES&630.25, 200 G & 13.8 & 1047 & 968 &     0.679 $\pm$     0.216&     0.339 $\pm$
0.140& -0.001 $\pm$     0.199& 8.70 $\pm$      2.06 \cr\hline
\end{tabular}\\
\tablefoottext{1}{OBS: observations}
\tablefoottext{2}{MHD-FR with average magnetic flux of 50, 100, 200\,G, respectively}
\end{table*}

\subsection{Line parameters\label{sect_res_lineparam}}
The spatial resolution of the HD simulation and the observations was matched
using the continuum intensity (Figs.~\ref{int_hist} and \ref{kernel_2d}),
whereas the spectral degradation was determined comparing average observed
and atlas profiles without using the simulations' spectra. The comparison of
the line parameters other than $I_c$ therefore can reveal possible mismatches
between the synthetic and observed spectra that can indicate shortcomings of
either simulations or observations (or both). Figure \ref{fig_stat_rest}
shows the histograms of the line asymmetry, FWHM, equivalent width, and line
depth for the HD-SPAT-SPEC and for the observations. We note that the size of the statistical sample for observations and
simulations differs by about one order of magnitude (cf.~the bottom row
in Table \ref{table1}). For all of the following histograms,
always the full FOV of the observations with several ten thousand profiles
was used, whereas the simulations only provide about 4000 profiles.

The line asymmetry (top left panel in Fig.~\ref{fig_stat_rest}) is seen
to be sensitive to the spatial resolution, as also found by PuB11. The shape
of the line asymmetry histogram in both observations and simulations changes
from a roughly Gaussian distribution at the spatial resolution of the
destretched GFPI data (second column, second row) to a broad, flatter
distribution with a tail towards negative asymmetries at the resolution of
the SP (top row). The line asymmetries in the observed POLIS spectra
show a much broader histogram with significantly larger absolute values than
the HD-SPAT-SPEC, which is also the case for most of the other line
parameters from POLIS in Fig.~\ref{fig_stat_rest} in one or both 630\,nm
lines. We suspect the presence of some spectral degradation effect that
varies along the slit such as a varying spectral focus as a reason, because
the average profiles matched very well (Fig.~\ref{av_prof}). The line depth
in the data set used is also somewhat lower than for other POLIS observations
\citep[cf.][their Fig.~6]{beck+etal2005b}. The line asymmetries of the
near-IR lines at 1565\,nm are smaller than those of all other lines because
of the smaller formation height ranges of the former.

The shapes of the histograms of the FWHM (top right panel) and
equivalent width (bottom left panel) are usually similar for the HD-SPAT-SPEC and the observations, but in some cases the histograms are
(slightly) displaced from each other. There is a slight tendency for the HD-SPAT-SPEC to have a lower FWHM, but there are also reverse cases. A
better match of the FWHM or equivalent width would require some fine-tuning
in their determination (e.g.~in the case of equivalent width measurements,
accounting for line blends) and in the synthesis setup (e.g.~abundances,
line broadening). This issue, however, should not have a critical impact on
the spatial distribution of the values caused by the thermodynamics. The line
depth (bottom-right panel) shows a very good match for all lines, except
for the \ion{Si}{i} line at 1082.7\,nm. All histograms share a common trend with increasing spatial resolution, i.e.~they become more asymmetric/skewed rather than only becoming broader. Figure \ref{fig_5576} shows the corresponding histograms of the \ion{Fe}{i} line at 557.6\,nm for completeness.

The histograms of the line-core velocity are shown in
Fig.~\ref{fig_stat_velo} (see Fig.~\ref{fig_5576} for 557.6\,nm). The
line-core velocity reflects the dynamics of the mass motions in the upper end
of the respective formation heights of the various spectral lines. The match
between the HD simulation and the observations is excellent, nearly as good as the one found for the continuum intensity distributions (Fig.~\ref{int_hist}) that we employed for the determination of the spatial degradation. For the SP data in 630.15\,nm and the near-IR data in 1082.7\,nm and 1565.2\,nm, the line-core velocity histograms for the observational data show a peculiar shape, with a central plateau at low velocities, a steep decline at intermediate velocities and a low probability of finding velocities above about $\pm$ 500 ms$^{-1}$. This shape turns out to be reproduced surprisingly well by the HD-SPAT-SPEC and appears also in other convection simulations \citep[e.g.][]{stein+nordlund1998,beeck+etal2012}.

 Table \ref{big_table} lists the average and rms values of some line 
parameters, respectively for the observations, the HD-FR, the HD-SPAT and the HD-SPAT-SPEC. We also included the parameters of the 630.25\,nm line in the
MHD simulations with increasing average magnetic flux (50\,--\,200 G) at full
resolution for comparison (lowermost three rows in Table
\ref{big_table}). 

The clearest trend in the observational values is the
increase in all rms fluctuations with increasing spatial resolution, yielding a factor of about 1.5 between, e.g.~POLIS and the SP in the rms
velocities (fourth column of Table~\ref{big_table}). The HD-FR shows rms velocities  \citep[compare the values with, e.g.][last column of their Table
    1]{asplund+etal2004} larger by an additional factor of about two in comparison to SP and GFPI observations. Whereas the spatial degradation of the simulations
always significantly decreases the corresponding rms fluctuations, the
spectral degradation has in some cases the opposite effect of increasing the
rms values, although only slightly (i.e.~spectral degradation has a weaker
impact than spatial degradation). The observed rms velocities are usually
matched in the HD-SPAT-SPEC to within about $\pm$\,50\,ms$^{-1}$. In
the case of the line-core velocity $v_{\rm core}$, the rms in the HD-SPAT-SPEC is lower than in the observations for all lines considered.

The line depth (sixth column of Table~\ref{big_table}) of the HD simulation is usually reduced only slightly by the spatial degradation and more strongly by the spectral degradation, whereas the line-core intensity  (seventh column of Table~\ref{big_table}) naturally shows a corresponding increase. The factor with the biggest impact on the changes in line depth and line-core intensity is the stray-light offset $\beta$ in the spectral degradation. The significant consequences of these changes in the line shape on the temperature stratifications retrieved by the inversion are discussed in Sect.~\ref{inv_results} later on.

The average line asymmetry (last-but-one column in
Table~\ref{big_table}) changes strongly with both the spatial or
spectral degradation of the simulations, even including a change of sign. In most cases, the change from positive to negative asymmetry (or reverse) by the
degradation is actually necessary to match the observed value (e.g.~for
557.6\,nm, for 630\,nm in the SP and GFPI data, and all near-IR
lines). Assuming the same temperature structure in granules and intergranular
lanes (IGLs), decelerating upflows (accelerating downflows) lead to a
positive (negative) line asymmetry. The spatial average of the line asymmetry measures an intensity-weighted relative area filling factor of granules and small-scale IGLs, which, after the latter are largely suppressed by the spatial degradation process, yields mainly positive values.

Accounting for spatial degradation reduces the equivalent width  (last column in Table~\ref{big_table}) of the spectral lines by $0.02-0.09$\,pm, depending on the line, except for GFPI 630.25 destretched, for which it increases it by 0.13 pm. The increase in magnetic flux in the MHD-FR reduces the equivalent width of the 630.25\,nm line by up to $\sim 0.3$\,pm (HD-200\,G). This effect is due to the warmer atmospheric temperature structure generally retrieved in the
presence of magnetic fields, that causes line-core weakening \citep[``{\em
    line-gap}'',
  e.g.][]{stellmacher+wiehr1971,chapman1977,hirzberger+wiehr2005}. The
corresponding decrease in equivalent width is the dominant effect of the
presence of magnetic fields on iron lines in the visible wavelength range,
even if they are also Zeeman-sensitive. At visible wavelengths, the Zeeman
broadening is not sufficiently strong to counter the indirect effect on
spectral lines due to the change in average temperature induced by magnetic
flux \citep{fabbian+etal2010}. The spectral degradation tends to reduce the
equivalent width by even larger amounts (up to several pm for the 1082.7 nm
line), because the spectral stray light offset $\beta$ (cf.~Table
\ref{tab_spec}) directly scales the line depth, and hence also the total area of the absorption profile.
\begin{figure}
\resizebox{8.8cm}{!}{\includegraphics{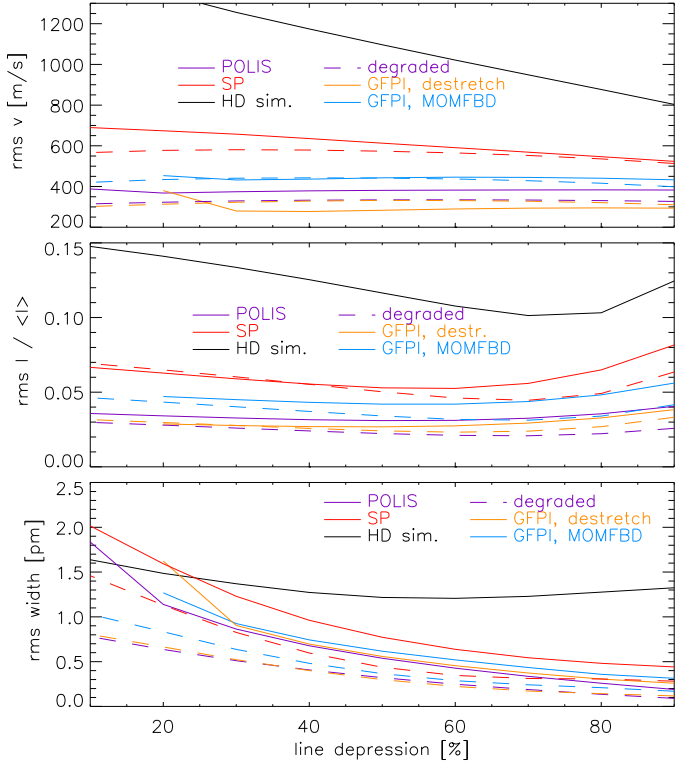}}
\caption{Rms fluctuations of the line properties at different line depression
  levels for the 630.25\,nm line in the observations (solid lines), in
  the HD-FR (solid black line) and in the HD-SPAT-SPEC (dashed lines). {\em Top}: bisector velocity. {\em Middle}: bisector intensity. {\em Bottom}: line width.\label{fig_bisec_rms}\label{bisec_int_rms}\label{bisec_length_rms}}
\end{figure}
\subsection{Line properties at different line depression levels\label{bisec_sec}}
The bisector is defined as the curve connecting the central positions between
left and right wing at different line depression levels throughout an
absorption line. We determined the bisector position and the line width at
ten different intensity levels throughout the absorption profiles. The variation of these properties in an individual spectral line profile traces the values of the corresponding thermodynamic stratification in the atmosphere and any possible gradients. A line depression of 0\,\% denotes the continuum level,
whereas 100\,\% denotes the line core. We used the average bisector position
at 50\,\% line depression as zero point reference for the conversion to
velocities.

\begin{figure}
\resizebox{8.8cm}{!}{\includegraphics{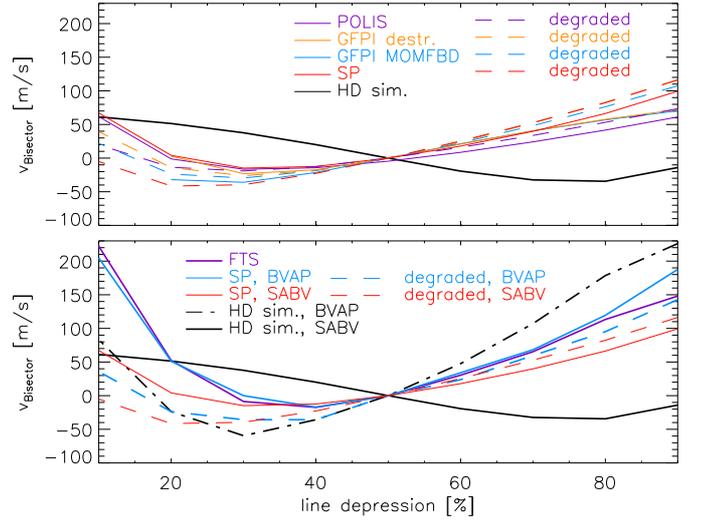}}
\caption{{\em Top panel}: spatially averaged bisector velocity (SABV) for the
  630.25\,nm line in the observations (solid lines), the HD-FR
  (solid black line) and the HD-SPAT-SPEC (dashed
    lines). {\em Bottom panel}: comparison between SABVs and bisector
  velocities of average profiles (BVAP) for the 630.25\,nm line. 
    Solid/dash-dotted black lines: SABV/BVAP in the HD-FR. Solid
    red/blue lines: SABV/BVAP for the SP data. Red/blue dashed lines: SABV/BVAP for the simulation degraded to SP resolution. The purple line gives the bisector in the FTS atlas profile. Positive velocities correspond to redshifts relative to the average velocity at 50\,\% line depression.\label{fig_bisec}}
\end{figure}

The bisector positions at different line depression levels trace gradients of
velocity with optical depth. For average spectra in QS granulation, the
bisectors usually yield a C-shape throughout the different levels of line
depression \citep[e.g.][]{brandt+schroeter1982,dravins1982,marquez+etal1996,puschmann+etal1999,hanslmeier+etal2000,asplund+etal2000,mikurda+etal2006,dravins2008,gray2010,delacruzrodriguez+etal2011}. The upper panel of Fig.~\ref{fig_bisec} shows the spatial average of the bisector positions, converted to the corresponding velocities, for the 630.25\,nm line in the observations and the HD simulation. A prominent difference is seen: the averaged bisector velocity of the HD-FR has a roughly inverse shape to that of all observations. The match between the velocity at different line depression levels in observed and synthetic spectra is, however, good for the
spatially and spectrally degraded HD simulation, except for the lines
at 557.6\,nm and 1082.7\,nm (Fig.~\ref{mean_bisc_pos_rest}). For 557.6\,nm,
the deviations at line depression levels below 40\,\% are presumably caused
by the weak blends in its line wing. Note that the spatial average of the bisector velocities does not reflect the dynamical range of the vertical velocity fluctuations, but rather the average net vertical flow speed.

To determine the reason for the inversion of the bisector shape between
observations and the HD-FR, we reversed the
order of the bisector determination and the spatial averaging by calculating
also bisectors for spatially averaged profiles. The lower panel of
Fig.~\ref{fig_bisec} compares the spatially averaged bisector
velocities (SABV) and the bisector velocities of average profiles (BVAP) for
the 630.25\,nm line. We included the SP data, the HD-FR, the HD simulation
spatially and spectrally degraded to the SP data, and the profile of the FTS
atlas. Applying the spatial averaging to the spectra before determining the
bisectors flips the shape of the bisector velocities with respect to the HD-FR
(compare the black solid and black dash-dotted lines in the lower panel of
Fig.~\ref{fig_bisec}). The similarity of the BVAP for the simulation (black
dash-dotted line) and the SABV for the HD simulation degraded to SP resolution
(red dashed line), or the SP data themselves (red solid line),
indicates that the change of the shape must be introduced by a small-scale
averaging between the full spatial sampling of the simulation (0\farcs13
$\sim$ 94\,km) and the resolution of the SP data (0\farcs62 $\sim$
450\,km). The BVAP of the HD-FR also matches the bisector shape of the
MOMFBD GFPI data (blue line in the upper panel of
Fig.~\ref{fig_bisec}), which puts an even stricter upper limit to the spatial
averaging effect of about 0\farcs3 $\sim$ 218\,km. The reversal of the shape
of the bisector should also correspond to the sign change of the line
asymmetry by the spatial degradation that was found before
(cf.~Sect.~\ref{sect_res_lineparam} and Table \ref{big_table}) and should be coupled to the relative area fraction of the bright granules and dark
intergranular lanes.
\begin{figure*}
\resizebox{17.6cm}{!}{\includegraphics{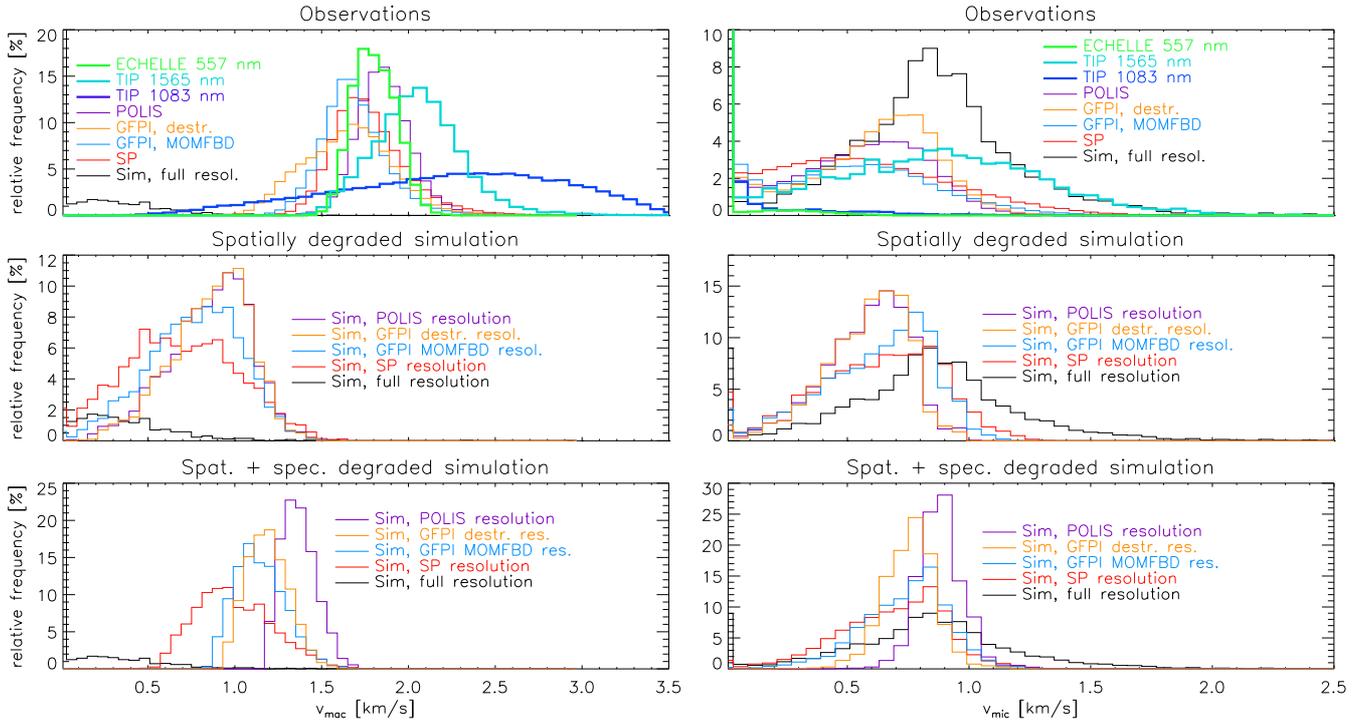}}
\caption{Histograms of the macroturbulent (left column) and microturbulent velocity (right column) in the inversions. {\em Top}: inversions of observations. {\em Middle}: the same for the HD-SPAT simulation. {\em Bottom}: the same for the HD-SPAT-SPEC. Thin black lines in each panel: HD-FR.\label{macvel_fig}}
\end{figure*}

Compared to the small-scale spatial average caused already
by the effective spatial resolution of the observations, any additional
large-scale averaging done in the BVAP process adds only minor
modifications. The SABV and BVAP of the SP data are similar (albeit not fully
identical), and the bisector from the FTS atlas profile matches the BVAP from
the SP data well for the 630.25\,nm line. These two facts are indicative of
two different subsequent averaging effects: on the one hand, the spatial
averaging caused by the limited spatial resolution of all of our observations
has already significantly changed the shape of the bisector velocity curve in
comparison to the HD-FR. On the other hand, averaging over the full FOV
(or a long time) leads to line profiles, and hence bisectors similar to FTS
regardless of the spatial resolution of the input spectra. This second
averaging step thus only produces minor variations of the bisector shape.

In contrast to the average position of the bisectors, the rms fluctuations of
the bisector velocities do reflect the vertical velocity fluctuations caused by the dynamical evolution of the solar atmosphere. For the SP observations both at 630.25\,nm (top panel of Fig.~\ref{fig_bisec_rms}) and at 630.15\,nm (Fig.~\ref{bisec_rms_rest}), and for the near-IR observations (Fig.~\ref{bisec_rms_rest}), an increase in the velocity rms is seen as one moves from the higher atmospheric levels towards the continuum formation layers in the
lower atmosphere, i.e.~for high bisector levels (line depression levels $<$
40\,\%). In the observed SP data at 630\,nm, the velocity rms decreases
monotonically with increasing line depression, like also in the HD-FR
for all spectral lines. Towards low line depression levels ($<40\,\%$) in
most lines, the degraded HD simulation shows lower rms fluctuations than the
observations, whereas the rms values are nearly identical near the
line-core. The bisector velocity rms for the 557.6\,nm line is shown in
Fig.~\ref{bisec_rms_5576} for completeness.
\begin{table*}
\caption{Average and rms\tablefootmark{1} of the macroturbulent and microturbulent velocity in kms$^{-1}$, and rms of the LOS velocity in the inversion.\label{macvel_table}}
\begin{tabular}{c|cccccccc}\hline\hline
\multicolumn{8}{c}{Observations}\cr\hline
 data & SP & GFPI, MOMFBD & GFPI, destr. & POLIS & TIP@1083\,nm & TIP@1565\,nm & 557.6\,nm\cr\hline
$v_{\rm mac}$ & 1.81 $\pm$ 0.23 & 1.72 $\pm$ 0.20 & 1.68  $\pm$ 0.25 & 1.88 $\pm$ 0.16 & 2.27 $\pm$ 0.65 & 2.11 $\pm$ 0.34 & 1.83 $\pm$ 0.12 \cr
$v_{\rm mic}$ & (0.56 $\pm$ 0.38)\tablefootmark{2} & (0.47 $\pm$ 0.36)\tablefootmark{2} & (0.57 $\pm$ 0.32)\tablefootmark{2} & (0.50 $\pm$ 0.31)\tablefootmark{2} & (0.06 $\pm$ 0.16)\tablefootmark{2} &(0.83 $\pm$ 0.48)\tablefootmark{2} & (0.25 $\pm$ 0.20)\tablefootmark{2}\cr
$v_{\rm mic} > 10$\,ms$^{-1}$ & 31\,\% & 34\,\% & 52\,\% & 29\,\% & 15\,\% & 42\,\% & $<$ 1\,\% \cr 
rms($v_{\rm LOS}$) & 744 & 443 &  362 &  444 &  497 &  500 & 319\cr\hline\hline
&\multicolumn{4}{c}{Spatially degraded HD simulation}&&& SIM\tablefootmark{3}\cr\hline
$v_{\rm mac}$  &  0.63 $\pm$ 0.39 & 0.78 $\pm$ 0.30 & 0.89 $\pm$ 0.23 & 0.88
$\pm$  0.24 && & (0.41 $\pm$ 0.32)\tablefootmark{4} \cr
$v_{\rm mic}$  &  0.55 $\pm$ 0.34 & 0.57 $\pm$ 0.31 & 0.58 $\pm$ 0.22 & 0.58 $\pm$ 0.22    && & 0.76 $\pm$ 0.45 \cr
rms($v_{\rm LOS}$)  & 686 & 534 & 405 & 416&& & 1058 \cr\hline\hline
&\multicolumn{4}{c}{Spatially and spectrally degraded HD simulation}\cr\hline
$v_{\rm mac}$  & 1.03 $\pm$ 0.22 &  1.18 $\pm$ 0.15 & 1.22 $\pm$ 0.12 & 1.40 $\pm$    0.10 \cr
$v_{\rm mic}$  & 0.68 $\pm$ 0.27 & 0.76 $\pm$ 0.19 & 0.79 $\pm$ 0.11 & 0.91 $\pm$ 0.10 \cr
rms($v_{\rm LOS}$)  & 685 & 529 & 400 & 410 \cr
\end{tabular}\\
\tablefoottext{1}{Numbers in parentheses indicate that the average was done only over the fraction of the FOV where the quantity was significant ($> 10$\,ms$^{-1}$)}
\tablefoottext{2}{only $v_{\rm mic} >$ 10\,ms$^{-1}$}
\tablefoottext{3}{630\,nm data at full resolution without any degradation}
\tablefoottext{4}{only $v_{\rm mac} >$ 10\,ms$^{-1}$, 10\,\% of the FOV}
\end{table*}

The decrease in the rms velocity fluctuations with height in the atmosphere
indicates that the velocity formation heights for most of the spectral lines
do not reach into the realm where propagating acoustic waves of increasing
amplitude dominate the atmosphere making the rms velocities increase again
strongly with height \citep[e.g.][]{beck+etal2009}. The spectral lines used
here therefore in general correspond to an height regime that is dominated by
the convective velocity distribution as far as the line-of-sight (LOS) velocities are concerned.

The bisector intensity rms shows a different height dependence (e.g.~middle panel of Fig.~\ref{bisec_int_rms}) than the bisector velocity rms. The relative (i.e.~normalized by the average intensity) rms fluctuations of the
intensity levels show increased fluctuations for line depression levels above
about 70\,\%. This indicates an increase in temperature fluctuations close to
the line core, as expected for propagating acoustic waves with a steepening
wave amplitude. This explanation stands at least for the HD simulation that is
free of magnetic flux, whereas in the observations it could also in principle be related to the presence of magnetic fields, because the latter lead to brightenings in the line core (cf.~\citeauthor{puschmann+etal2007} \citeyear{puschmann+etal2007}, or Fig.~\ref{fig_1}, middle row: the brightenings in the lower left corner of the FOV for the GFPI data are of a magnetic origin (see PuB11)). The rms of the intensity in the observations generally exceeds the fluctuations in the HD-SPAT-SPEC.

The rms fluctuations in the line width at different line depression levels for
630.25\,nm  are shown in the bottom panel of
Fig.~\ref{bisec_length_rms}. In this case, the rms fluctuations start to
increase when reaching about 60\,\% in line depression for the 630\,nm lines
(as well as for the 557.6\,nm line, Fig.~\ref{bisec_rms_5576}) in the HD-FR,
whereas all observations and the HD-SPAT-SPEC show a monotonic decrease in the
rms with increasing line depression. The fluctuations of all parameters in the
spatially degraded HD simulation generally are smaller than those in the observations besides for the low-forming 1565\,nm lines, where the HD-SPAT-SPEC match the observations nearly perfectly (Fig.~\ref{bisec_rms_rest}). The average values of the bisector intensity and the line width at different line depression levels are shown in Figs.~\ref{bisec_int} and \ref{bisec_length} for completeness.
\subsection{Inversion results \label{inv_results}}
In the inversion, the macroturbulent velocity $v_{\rm mac}$ is used to
convolve the synthetic profiles with a Gaussian of variable width to match the
observed spectra. The microturbulent velocity $v_{\rm mic}$ is one of the
contributions to the line width in a Lorentzian function that is used in the
integration of the radiative transfer equation \citep[see, e.g.][]{rutten2003}. The value of $v_{\rm mic}$ could be specified to vary on the optical depth scale in the SIR code, but we set it to be constant at all $\tau$ values. The two ad-hoc empirical parameters $v_{\rm mac}$ and $v_{\rm mic}$ encompass two different physical effects when observed spectra are inverted: the ``real'' macro/microturbulence caused by unresolved macro/microscopic velocities in the solar atmosphere, and the spectral broadening caused by the characteristics of the respective spectrometer. While the former component should scale with the spatial resolution, the latter should be a constant characteristic of the spectrometer that in some inversion codes can be included by providing a measurement of the PSF such as those in Sect.~\ref{spec_deg}. In the case of the simulations without any degradation, both effects are absent; for the spatially degraded simulations, only the broadening by unresolved velocities is included.

The top row of Fig.~\ref{macvel_fig} shows the histograms of the macroturbulent (left column) and microturbulent (right column) velocities derived by applying the SIR code for the inversion of the different observed spectra. Only one value per wavelength range was derived. Even when multiple spectral lines were present in a given range, they were in fact inverted simultaneously. The middle and bottom row of Fig.~\ref{macvel_fig} show the macroturbulent and microturbulent velocities retrieved from inverting the HD-SPAT and HD-SPAT-SPEC, respectively. The values from inverting the HD-FR are overplotted with black lines in each panel as reference. 
\paragraph{Macroturbulent velocity} There is an offset in macroturbulent velocity between the observations at 630\,nm (most $v_{\rm mac} < 2$\,kms$^{-1}$) and in the near-IR (most $v_{\rm mac} > 2$\,kms$^{-1}$) that is presumably caused by the lower spectral sampling and/or resolution of the near-IR observations with TIP (left top panel of Fig.~\ref{macvel_fig}). The distribution of $v_{\rm mac}$ for the 1082.7\,nm line is significantly broader than those obtained for any of the other observations. This was found to result from using only two nodes in the temperature stratification. An inversion of the observation in this line using three nodes yielded a distribution of $v_{\rm mac}$ that was similar to that of all other observations, with a much smaller width and a roughly Gaussian shape. The \ion{Si}{i} line at 1082.7\,nm has an extended formation height and is sensitive to NLTE effects \citep[cf.][and references therein]{shchukina+etal2012}.
\begin{figure*}
\resizebox{17.6cm}{!}{\includegraphics{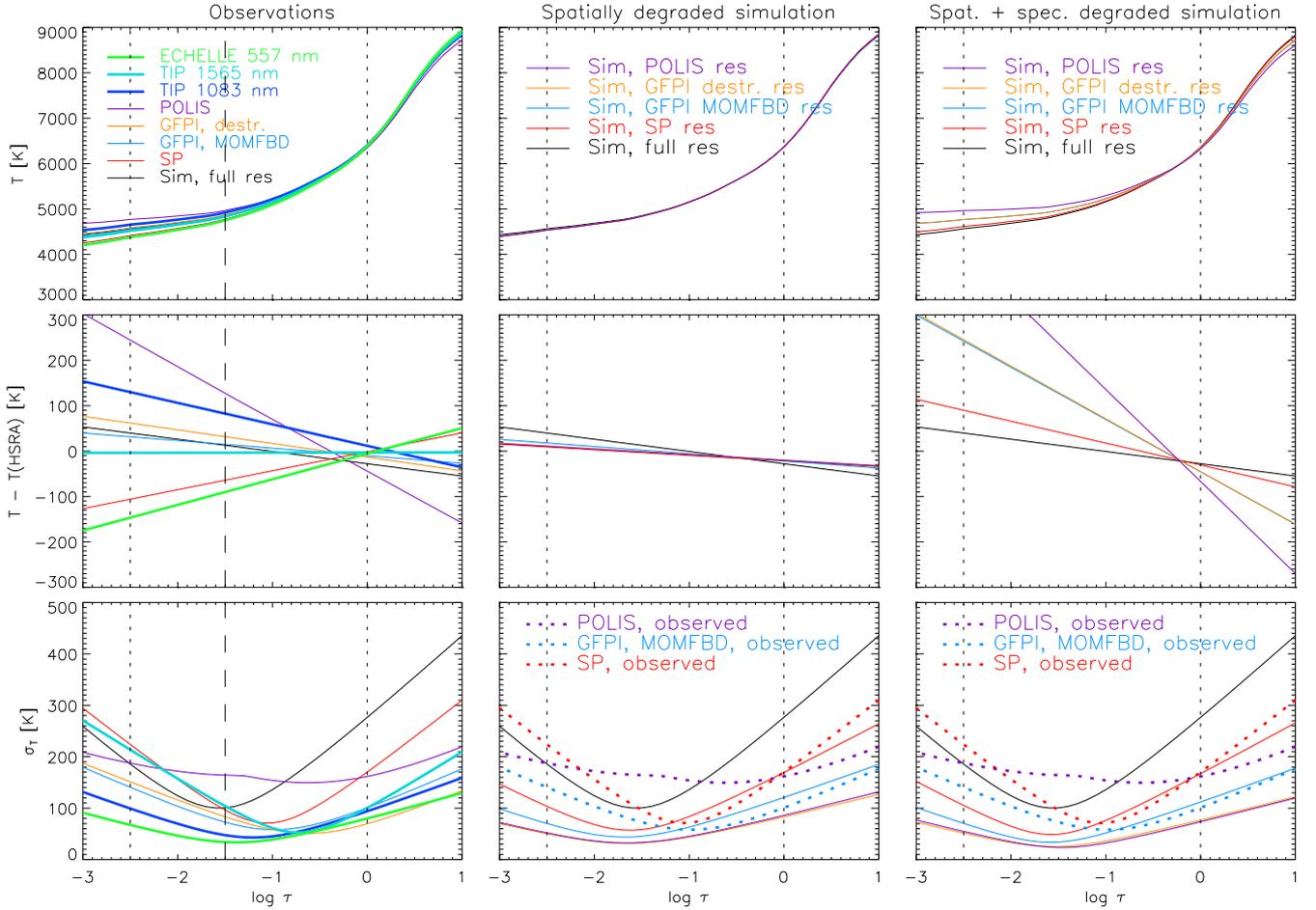}}
\caption{Temperature stratifications ({\em top panel}), difference to the HSRA
  model ({\em middle panel}), and rms fluctuations of temperature ({\em bottom panel}) in
  the inversions. {\em Left to right}: observations, HD-SPAT and HD-SPAT-SPEC. The dotted vertical lines denote the approximate formation height range of the 630\,nm lines, the dashed vertical line the upper end of the range for near-IR lines. Other labels are identical to Fig.~\ref{macvel_fig}. \label{tempinv_fig}}
\end{figure*}

For all 630\,nm observations, a slight trend with spatial resolution can be seen, with POLIS yielding the highest and the GFPI the lowest average $v_{\rm mac}$ (1.7\,--\,1.9\,kms$^{-1}$, cf.~Table \ref{macvel_table}). For the observational data, the spatial structuring in 2D maps of $v_{\rm mac}$ roughly mirrored an inverse granulation pattern, with increased $v_{\rm mac}$ in inter-granular lanes, especially on all locations where polarization signal was also present. Such a characteristic resulting pattern is likely related to the one-component inversion setup, where the missing freedom in the magnetic fill factor forced the inversion code to emulate part of the Zeeman line broadening in intensity by an increased macroturbulence. The HD-FR exhibited a negligible macroturbulence, with $v_{\rm   mac} < 10$\,ms$^{-1}$ on 90\,\% of the pixels and an average $v_{\rm mac}$ of 0.4\,kms$^{-1}$ on the remaining 10\,\% of the FOV. The few pixels with increased $v_{\rm mac}$ in the HD-FR were almost exclusively located inside inter-granular lanes.

The macroturbulent velocity introduced into the synthetic spectra by the
effect of spatial degradation alone (left middle panel of
Fig.~\ref{macvel_fig}) has average values increasing from 0.6 to
0.9\,kms$^{-1}$ with decreasing spatial resolution (Table \ref{macvel_table}). The value of $v_{\rm mac}$ was significant on all pixels
in the FOV of the HD-SPAT. The macroturbulent velocity
thus becomes significant between the spatial sampling of the HD simulation of
0\farcs13 and the highest spatial resolution in the observations of
0\farcs3. The spatial structuring of $v_{\rm mac}$ in the HD-SPAT resembled an inverse granulation pattern, but only remotely. 

The inversion of the HD-SPAT-SPEC spectra (left bottom panel of Fig.~\ref{macvel_fig}) yielded an average $v_{\rm mac}$ lower by about 0.4\,--\,0.8\,kms$^{-1}$ than for the observations. The final values from the inversion of the HD simulation agree well with the quadratic addition of the two contributions from the spatial and spectral degradation, indicating that the macroturbulent velocity retrieved by the inversion is a rather ``robust'' quantity that reliably captures the different broadening contributions. As an example, for the HD simulation degraded to SP resolution one has $<v_{\rm mac}> = 1.03$\,kms$^{-1} \sim \sqrt{0.83^2+0.63^2} = 1.04$\,kms$^{-1}$ (Tables \ref{tab_spec} and \ref{macvel_table}), which nevertheless falls short of the value in the observations (1.81\,kms$^{-1}$).
\paragraph{Microturbulent velocity} The microturbulent velocity $v_{\rm
  mic}$ in the HD simulation and the observations behaved roughly opposite to $v_{\rm  mac}$. The value of $v_{\rm mic}$ stayed below 10 ms$^{-1}$ on a large
fraction of 50\,--\,85\,\% of the FOV for all observations (Table \ref{macvel_table}, top part), contrary to $v_{\rm mac}$. It was instead significant on every pixel in the inversions of the synthetic HD-FR. The average microturbulent velocity for simulations with any type of degradation exceeded the value in the observation -- again contrary to the behavior of macroturbulent velocity. The spatial maps of $v_{\rm mic}$ for the HD-FR spectra exhibited a granular pattern, whereas those of the observations resembled $v_{\rm mac}$ of the HD-FR, i.e.~for the observational data $v_{\rm mic}$ was only significant in a few points in inter-granular lanes. The HD-SPAT yielded a nearly identical average $v_{\rm mic}$ of about 0.57\,kms$^{-1}$ regardless of the amount of spatial degradation. After the additional spectral degradation, $v_{\rm mic}$ increases with decreasing spatial resolution (Table \ref{macvel_table}, bottom part) as for $v_{\rm mac}$.

Despite the effort in matching synthetic and observed spectra spatially
  and spectrally, the inversion with SIR retrieves a quite different behavior
  with respect to $v_{\rm mac}$ and $v_{\rm mic}$. The reason most likely is
  the usage of the Uns\"old approximation by LILIA in the synthesis, whereas SIR used the ABO formalism in the inversion of the observed or synthetic spectra. In any case, this should not have had a significant influence on the temperature fluctuations retrieved by the inversion.
\paragraph{Temperature} The last indicators of thermodynamics provided by the inversion are the temperature stratifications $T$, and -- more relevant for the current study -- the rms fluctuations $\sigma_T (\tau)$ of the temperature at a given optical depth. Figure \ref{tempinv_fig} shows the average temperature stratifications (top), their deviation from the HSRA model (middle), and their rms fluctuations (bottom) vs.~optical depth for the observations (left column), the HD-SPAT (middle column), and the HD-SPAT-SPEC (right column). 

The average temperature stratifications retrieved from the inversion of the observations differ somewhat (top left panel), but overlap between them or with the HSRA model within $\lessapprox \pm 100$\,K in the range of log\,$\tau = 0$ to $-1$ (left middle panel of Fig.~\ref{tempinv_fig}). For log\,$\tau <
-1$, they show a considerably larger scatter between the different instruments
and increasing temperature deviations. The slope (from lower to higher optical
depth) of the deviation from the HSRA changes from being positive for the SP
data to slightly negative for the GFPI data, and finally becomes strongly
negative for the POLIS data. The temperature stratification retrieved by
inverting the 630\,nm spectra from the HD-FR would
correspond roughly to the average of all other stratifications derived from the observations. The deviation (from the HSRA model) of the average temperature stratification retrieved from inverting the HD-FR shows a slightly positive slope, with differences of $\lesssim \pm 50$\,K.

We note that the trend in the slope of the deviation from HSRA (SP $<$ GFPI
$<$ POLIS) in the case of the inversion of the observational data does not
follow the spatial resolution trend of the latter (GFPI $<$ SP $<$ POLIS). A
different suspect instead clearly comes to mind for which the trend is the
same as for the slopes: namely, the stray light offset inside the respective
spectrometers (cf.~Table \ref{tab_spec}). Whereas $v_{\rm mic}$ and $v_{\rm
  mac}$ can account for the spectral broadening in each instrument, the
constant intensity offset $\beta$ was not considered in the inversion. The
hypothesis that the slope is related to the spectral stray light offset is
further confirmed when considering the inversions of the HD simulation
degraded to GFPI resolution, where we have two cases with different spatial
but identical spectral resolution. The deviations from HSRA after degrading
the HD simulation only spatially to the spatial resolution of the MOMFBD or to the destretched GFPI spectra stay small and both slopes are almost
identical (central panel, lines are overlapping) even if the spatial
resolution is clearly different (Table~\ref{table1} and
Fig.~\ref{kernel_2d}). After the additional application of the spectral
  degradation to the spatially degraded HD simulation spectra, the slope of
  the deviation from HSRA increases significantly for the synthetic data at
  GFPI resolution (middle right panel), lying between that of the HD simulation at SP or POLIS resolution, but with again little difference between the two different spatial resolutions of the GFPI data (orange and blue line overlapping again). This indicates that, contrary to the spatial resolution, the spectral resolution has a strong effect on the deviation from HSRA, likely due to stray light effects unaccounted for in the inversions.

The rms fluctuations of $T$ are shown in the bottom row of
Fig.~\ref{tempinv_fig}. For the observations (bottom left panel), the
values range from 100 to 200\,K at log $\tau = 0$, pass through a minimum of
below 100\,K rms between log $\tau \sim -0.7$ and $-1.2$, and tend to increase
to up to $\sim 300$\,K at log $\tau \lesssim -2.5$. This agrees with the
results obtained by \citet{puschmann+etal2003,puschmann+etal2005}, who found a
minimum temperature contrast between granules and IGLs at $\log \tau =
-1$. The HD-FR yields about 100\,K larger fluctuations in the lower atmosphere
up to log $\tau = -1$, but lower rms values than for instance the SP data
above that height. The shape of the rms curve for the HD simulation is similar
to that of the inversion of, e.g.~the SP data, besides an offset in log $\tau$
of about 0.75 (compare the location of minimal rms in both curves). This
offset could be caused by different values of gas (or electron) densities
because for instance in the inversion with SIR the electron density at log
$\tau$ = +1.4 was fixed to the value in the HSRA as a boundary
condition. After the spatial degradation of the HD simulation, the rms
temperature fluctuations retrieved from the inversion with SIR match the
observations in the lower photosphere from log $\tau = 0$ to $-1$ well
(compare the red/blue solid and red/blue dotted curves in the bottom middle panel of Fig.~\ref{tempinv_fig}), whereas they are smaller than the corresponding observed rms values for higher layers. The subsequent spectral degradation has no additional visible effect on the rms temperature fluctuations (bottom right panel), while it significantly changes the average temperature stratification, hence producing the difference to the HSRA (middle right panel).

The scatter of the retrieved average temperature stratifications from the HD-SPAT-SPEC becomes comparable to
that between the different observations (top left and top right
  panels). The deviations between the average temperature stratification from
the inversion of the HD-SPAT-SPEC and the HSRA (middle right panel) follow the same systematic trend as in the case of the observations (middle left panel), i.e.~a decrease in the slope from SP to POLIS, but in this case starting already from a negative slope for the SP data. That the HD-SPAT-SPEC show the same trend as the observations -- while the HD-SPAT (central panel) shows a basically zero slope  -- again confirms that the spectral degradation, and there mainly the stray light offset, are responsible for the variation in the derived temperature stratifications. The stray light offset changes mainly the line depth, and therefore compresses or expands the attributed optical depth scale in the inversion. 

The deviations from the HSRA (or any other semi-empirical reference model)
would be a concern for an abundance determination, but in the present study we
are only interested in the behavior of observations and simulations relative
to each other when using the same analysis approach for both. In particular,
we focus on the rms fluctuations rather than on the average values. In that
respect, the rms temperature fluctuations as derived from the degraded HD
simulation do not exceed the observed properties. Rather, they match them
fairly well at layers of formation of photospheric spectral lines commonly
used for abundance derivations, and actually are smaller than the observed ones at high layers.
\subsection{Comparison of temperature fluctuations to other studies}
\begin{figure}
\resizebox{8.8cm}{!}{\includegraphics{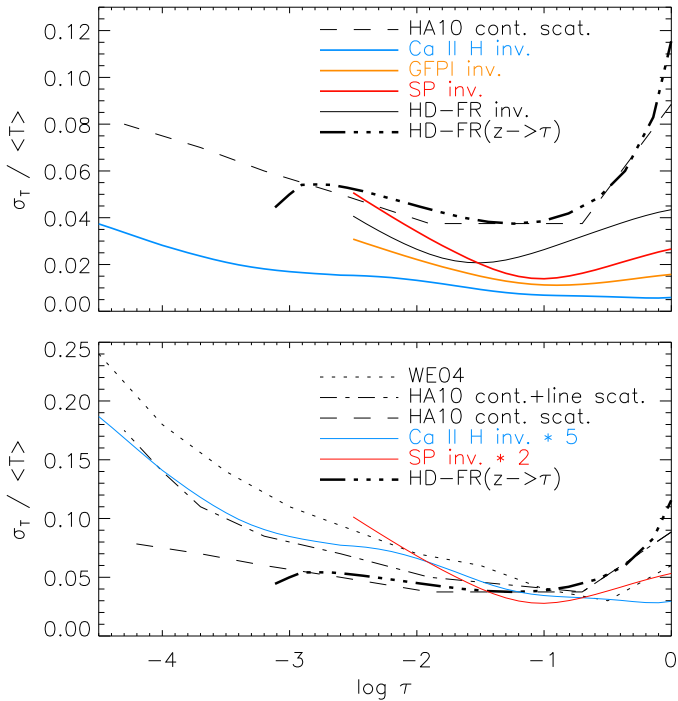}}
\caption{Relative temperature fluctuations vs.~optical depth. {\em Top
    panel}: inversion results of SP (red line), GFPI (orange
  line) and \ion{Ca}{ii} H data (blue line), and of the HD-FR simulation
  spectra (black-solid line). Overplotted are the relative temperature
  fluctuations in the HD-FR simulation box (black dashed-triple-dotted line)
  and the curve given in \citet[][HA10]{hayek+etal2010} for a simulation with
  continuum scattering (black-dashed line). {\em Bottom panel}:  inversion
  results of SP/\ion{Ca}{ii} H multiplied by two/five (red/blue lines), and the
  values from the simulations of \citet[][black-dotted
  line]{wedemeyer+etal2004}, of HA10 with continuum scattering (black-dashed
  line) and of HA10 with continuum and line scattering (black-dash-dotted
  line).\label{rmst_fig}} 
\end{figure}

The temperature fluctuations retrieved by the inversion of the degraded HD
  spectra do not exceed the observed values. To investigate whether this could
  be caused by a too small amplitude of fluctuations in our simulations, we
  compared the temperature fluctuations in our data (simulations and
  observations), with the values provided for other simulations (Fig.~\ref{rmst_fig}). All curves obtained from an inversion of
  630\,nm spectra were clipped at log $\tau = -2.5$ because the formation
  height of the lines does not cover higher layers. To extend the height range covered from the
  observational side, we included the LTE inversion results of \ion{Ca}{ii} H
  spectra from \citet{beck+etal2013,beck+etal2013a}. 

We read off the results given in
  \citet[][WE04; their Fig.~9]{wedemeyer+etal2004} and \citet[][HA10; their
  Fig.~10]{hayek+etal2010} for comparison because they correspond to different
  treatments of the radiative transfer in the simulations, i.e.~gray atmosphere
  in LTE (WE04) and LTE with additional continuum and line scattering
  (HA10). Because these simulations were not spatially degraded, we
  overplotted the results for the inversion of the SP (\ion{Ca}{ii} H) data
  multiplied by two (five) in the lower panel of Fig.~\ref{rmst_fig}. These
  factors roughly correspond to the ratio of continuum contrast between
  observations and simulations, e.g.~for the SP (\ion{Ca}{ii} H) one has an
  observed rms $I_c = 7.2$\,\% (3.5\,\%), whereas simulations predict about
  15\,\% \citep[$>$20\,\%;][]{hirzberger+etal2010}. We used the tabulated
  $z-\tau$ relation in the HSRA model to convert curves vs.~geometrical height
  (HA10, HD-FR($z\rightarrow \tau$)\,) to an optical depth scale. This
  approach can slightly compress or stretch the curves, when the true density
  stratification differs from the one of the HSRA. 

  The best match between an observed and a simulation's curve is
  obtained for the temperature rms derived from \ion{Ca}{ii} H, after scaling it up for the spatial resolution, and the results of HA10 including continuum and line scattering (lower panel of Fig.~\ref{rmst_fig}). The rms fluctuations in WE04 exceed all others above $\log \tau = -1$. The
  temperature  fluctuations in the \texttt{Stagger} code simulations that we
  used closely match the HA10 case with only continuum scattering included. The rms fluctuations of all simulations are
  comparable up to about $\log \tau = -2$ with relative fluctuations between 5
  and 7.5\,\% at this optical depth, but increasingly deviate in higher layers
  with differences of up to 15\,\% at $\log \tau = -4$ ($\equiv$ up to 600\,K
  difference in rms). From the comparison, we would exclude that our  \texttt{Stagger} simulations
  are somehow biased towards an extraordinary low or high level of fluctuations in the photosphere up to $\log \tau = -2$, which is about the maximal height range where the spectral lines used here are fully reliable.
\section{Summary and Discussion\label{sec_disc}}
Numerical (M)HD simulations are increasingly used in the
analysis or interpretation of observational data of the Sun
\citep[e.g.][]{khomenko+etal2005,shelyag+etal2007,carroll+kopf2008}, and in
particular also for the derivation of abundances
\citep[][]{asplund+etal2004,caffau+etal2008,pereira+etal2009a,fabbian+etal2010,fabbian+etal2012}.
They are capable of predicting phenomena such as small-scale vortices
\citep[e.g.][]{stein+nordlund1998,moll+etal2011,kitiashvili+etal2011} that
have been confirmed by observations only later on
\citep[][]{bonet+etal2008,steiner+etal2010,steiner+rezaei2012}. The claim
that a simulation is ``realistic'' is, however, often made solely based on
the physics included in the numerical calculation, and not necessarily
because the simulations were proven to correspond to the ``real'' Sun as far
as it can be determined from observations. Several different tests of the
realism of simulations were performed in the past, but mainly concentrating
on a few selected (average) parameters such as the continuum contrast
\citep{wedemeyer+etal2009,hirzberger+etal2010}, its CLV
\citep{koesterke+etal2008}, average profiles
\citep{asplund+etal2000,stein+nordlund2000,allendeprieto+etal2004} and their
CLV \citep{asplund+etal2004}, or average temperature stratifications. Only
recently, also spatially resolved properties of individual line profiles were
used for comparisons of simulations and observations
\citep[e.g.][]{pereira+etal2009b,pereira+etal2009}. Additional cross-checks
have recently been done between different simulation codes
\citep[e.g.][]{beeck+etal2012}, rather than between simulations and
observations. Some of the plots in \citet{beeck+etal2012} can, however, be
directly compared to those in the present study. We only point out one
feature of the temperature rms vs.~$\log \tau$ (their Fig.~12), namely the
location of minimum rms at about $\log\tau \sim -0.7$, matching the result in
our inversions of the observed spectra ($\log\tau \sim -1$) well, with a
similar result also in the inversions of \citet{puschmann+etal2005}.

All types of comparisons between simulations and observations usually share the same problem: the spatial resolution of the observations is not as good as that of the simulations, and thus it can only be shown that the simulations can be degraded sufficiently to match the observations, without the possibility to
investigate if the original simulation results match the ``real'' Sun
\citep[cf.~also][]{danilovic+etal2008}. Our present study demonstrated that
to do this final step would require observational data with about three to four times better spatial resolution than currently available. Such data will be
available soon with the upcoming telescope projects of the 1.5-m class such
as the GREGOR solar telescope \citep[e.g.][]{schmidt+etal2012a,schmidt+etal2012} with the GREGOR Fabry-Per{\'o}t Interferometer \citep[e.g.][]{puschmann+etal2012a,puschmann+etal2012c,puschmann+etal2012b} and
the New Solar Telescope \citep{denker+etal2006,cao+etal2010}, or in the
future with next generation solar telescopes of the 4-m class such as ATST
\citep{wagner+etal2008,rimmele+etal2010} or EST \citep{collados+etal2010}.

Our investigation shows that using 3D convection simulations as reference
allows one to derive an estimate of the spatial point spread function of
instruments and telescopes that can be used for an accurate stray light
correction and/or for a spatial deconvolution
\citep[BE11,][]{loefdahl+scharmer2012}. The good match to a directly measured
instrumental PSF (Fig.~\ref{kernel_polis_fig}) implies that the use of the
simulations for this purpose is well justified. The importance of knowing the
instrumental PSF should not be underestimated. The inversion of the HD-SPAT shows that the spatial smearing introduces a significant
broadening of spectral lines already at a resolution of 0\farcs3. Taking the
PSF explicitly into account in the analysis of data will thus yield more
accurate solar properties in for instance an inversion of spectra not only
for thermodynamic parameters, but also for magnetic field properties. The
PSF is commonly taken into account to correct for unpolarized stray
light only, but naturally applies as well to polarized stray light,
where it has a more profound impact on retrieved solar parameters, e.g.~for
all spectral lines in the weak-field limit where the polarization amplitude is
proportional to the total magnetic flux, or wherever spatial resolution is critical because of solar fine-structure such as in the penumbrae of sunspots \citep[][]{vannoort2012}.

The comparison of the HD-SPAT-SPEC data to the observations yielded a fair to
very good match for most line parameters. The HD-FR differs usually only by an
increased amplitude of the quantities, with the exception of the average
bisector position (Fig.~\ref{fig_bisec}). For this parameter, we find that its
average  values at different line depression levels trace a curve that for the
HD-FR has a roughly reverse shape compared to the one for the spatially
degraded HD simulation and the ones corresponding to the different
observational data. This change thus has to happen at a spatial scale between
0\farcs13 and 0\farcs3, close to the photon mean free-path length in the solar
photosphere. Given the corresponding thermalization length, one does not
  expect significant spatial temperature variations in the photosphere on
  spatial scales much smaller than 0\farcs13 ($\sim 100$\,km), but this argument does not hold in the same way for spatial velocity variations. The presence of Doppler shifts also enlarges the thermalization length \citep{skartlien2000}. The characteristic properties of the average bisector come from an intensity-weighted spatial average of the corresponding velocities. The difference in shape between the bisector derived from the HD-FR and those derived from any observed spectra then presumably is caused by both mass flows and spatial temperature variations that remain unresolved in the observations, while the spatial temperature variations should be partially resolved or close to being resolved in our best observations.

The bisector flips to the observed shape with any large-scale averaging. Our finding therefore should have no impact on abundance determinations from
averaged spectra. It, however, raises the warning that, since the
characteristic properties of spatially averaged spectral line profiles derived
from observations do not necessarily reflect those of spatially resolved
profiles, achieving a match with the former is not fully sufficient to prove the realism of simulations. Many line parameters such as bisectors depend on the exact line shape caused by gradients of thermodynamic parameters. A spatial averaging can destroy the (non-linear) dependence of the profile shape on the thermodynamic parameters. 

We point one peculiarity of the often used atlas spectra such as the FTS: the smearing of these data due to spatial averaging  and the corresponding amount of macroturbulent broadening are practically unknown. Spatial smearing introduces a significant line broadening of $v_{\rm mac} \sim 1$\,kms$^{-1}$ at a resolution of
0\farcs3, as the inversion of the HD-SPAT showed. In the case of spatially resolved observations, one can estimate the characteristic macroturbulent line broadening of the data in the resolved and averaged spectra. For the atlas spectra, this estimate cannot be done post-facto. The commonly adopted implicit assumption that the broadening by macroturbulent velocities levels off to a constant value at some degree of averaging should be investigated in more detail.

The simulation box covered a height range up to  more than 400\,km above optical depth unity in the solar  atmosphere, and the spectral lines used in this study -- besides the \ion{Si}{i} line at 1082.7\,nm that is not fully reproduced -- form in the  low to mid photosphere. The variation of quantities with height, like for  instance, a monotonic decrease in rms velocities, indicates that the values retrieved from the spectral lines still belong to an atmospheric regime whose
dynamics are governed by the convective energy transport. Any possible flaw of
the numerical simulations because of the approximations used (namely, LTE and opacity binning treatment) for the radiative energy transfer and losses is thus less relevant in these layers because the directed mass motions control the energy balance. A similar study covering also (low) chromospheric layers, e.g.~up to 1\,Mm height, should however not be too difficult to achieve in the near future \citep[cf.][]{wedemeyer+etal2004,leenaarts+etal2007,leenaarts+etal2009,wedemeyer+carlsson2011}. 

The inversion of the HD-SPAT yielded a small ($\lesssim \pm 50$\,K) spread in the resulting average temperature stratifications even if the spatial resolution varied between 0\farcs3 and about 1$^{\prime\prime}$ (central panel of Fig.~\ref{tempinv_fig}). The spectral degradation on the other hand led to a strong variation in the retrieved temperature stratifications, with a clear scaling by the constant wavelength-independent stray light offset  $\beta$ inside the spectrometer. This implies that for an accurate derivation of temperatures from observed spectra the stray light offset has to be corrected before an inversion. On the positive side, the stray light offset $\beta$ inside the SP spectrograph was found to be below 3\,\% \citep[see also][]{lites+etal2013a}, which makes these data and data of similar spectral quality well suited for a derivation of temperature stratifications in the solar photosphere \citep[e.g.][]{socasnavarro2011}.

The size of the statistical samples of simulations and
observations in our case are not identical. The FOV of the simulations of (6
Mm)$^2$ covers only a small fraction of that of the observational data
used. The close match between characteristic quantities indicates that any
statistical effect should, however, not have a significant impact. The line
properties change drastically with the spatial resolution, but in the same
way in degraded simulations and observations. An inclusion of more simulation
snapshots therefore is likely to confirm the close match
between the statistics of line parameters from observations and degraded
simulations. As a test, we compared the line profile
resulting from averaging over one or over 21 snapshots covering the whole HD
simulation series. The corresponding average profiles differed only slightly,
maintaining similar differences towards FTS or the observed spectra as those
seen in Fig.~\ref{av_prof}.
\section{Conclusions\label{sec_concl}}
We find no indications that the thermodynamic fluctuations in a
state-of-the-art numerical hydrodynamical simulation exceed those in
observations of the solar photosphere when the simulation is spatially
degraded to a resolution of 0\farcs3 or worse. For a final proof that the
initial properties of the simulation at full resolution do not exceed solar
surface properties, observations that resolve the photon mean free-path length
($< 0\farcs1$) are required. Solar telescopes with free apertures of about 1.5\,m and next-generation solar telescopes of the 4-m class are designed to achieve step by step this objective.
\begin{acknowledgements}
The VTT is operated by the Kiepenheuer-Institut f\"ur Sonnenphysik (KIS) at
the Spanish Observatorio del Teide of the Instituto de Astrof\'{\i}sica de
Canarias (IAC). The POLIS instrument has been a joint development of the High
Altitude Observatory (Boulder, USA) and the KIS. {\em Hinode} is a Japanese
mission developed and launched by ISAS/JAXA, collaborating with NAOJ as a
domestic partner, NASA and STFC (UK) as international partners. Scientific
operation of the Hinode mission is conducted by the Hinode science team
organized at ISAS/JAXA. This team mainly consists of scientists from
institutes in the partner countries. Support for the post-launch operation is
provided by JAXA and NAOJ (Japan), STFC (U.K.), NASA, ESA, and NSC
(Norway). C.B. acknowledges partial support by the  Spanish Ministerio de
Ciencia e Innovaci\'on through project AYA 2010-18029. R.R. acknowledges financial support by the DFG grant RE 3281/1-1. DF gratefully acknowledges financial support by the European Commission through the SOLAIRE Network (MTRN-CT-2006-035484) and by the Programa de Acceso a Grandes Instalaciones Cient\'ificas financed by the Spanish Ministerio de Ciencia e Innovaci\'on. The latter is also thanked by DF and FMI for providing funds through the related projects AYA2007-66502, CSD2007-00050, AYA2007-63881 and AYA2011-24808. The simulations were possible thanks to time awarded on the MareNostrum (BSC/CNS, Spain), the Danish Centre for Scientific Computing (DCSC-KU, Denmark), LaPalma (IAC/RES, Spain) and of the HRLS/DEISA (Germany) supercomputer installations. We thank A. Asensio Ramos and A. Lagg for their helpful comments on the present article.
\end{acknowledgements}
\bibliographystyle{aa}
\bibliography{references_luis_mod}
\Online
\begin{appendix}
\section{Line properties at different line depression levels\label{appa}}

\subsection{Average values}
\begin{figure}
\resizebox{8.8cm}{!}{\includegraphics{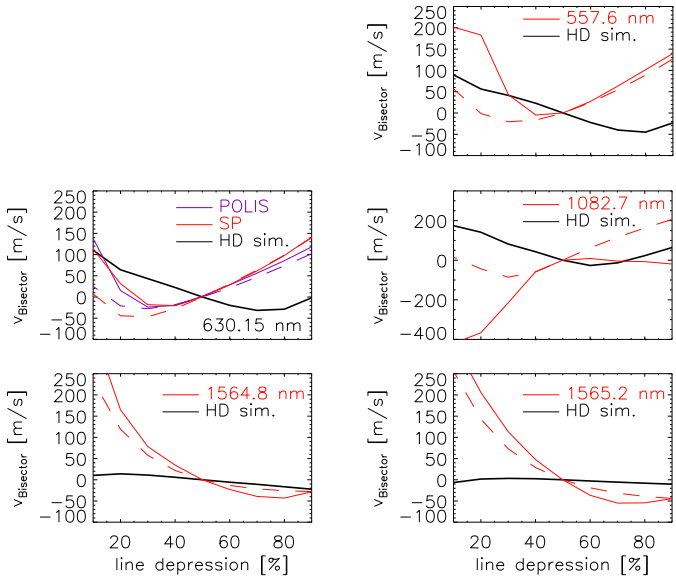}}
\caption{Average velocities at different line depression levels for all lines
  but 630.25\,nm in the observations (solid lines), the HD-SPAT-SPEC (dashed lines), and the HD-FR (solid black line).\label{mean_bisc_pos_rest}}
\end{figure}

\begin{figure}
\resizebox{8.8cm}{!}{\includegraphics{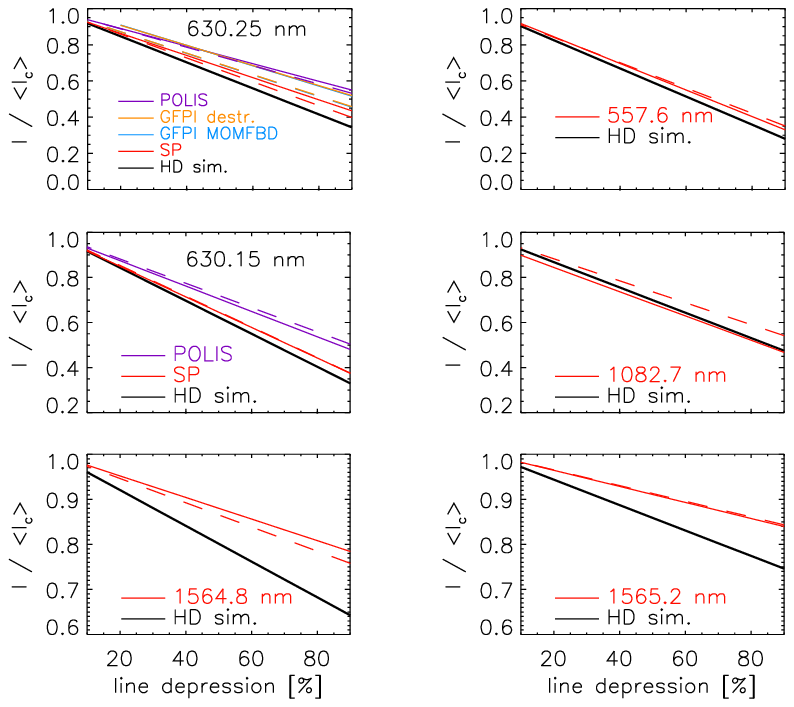}}
\caption{Average intensity at different line depression levels in the
  observations (solid lines), the HD-SPAT-SPEC (dashed
    lines), and the HD-FR (solid black line). \label{bisec_int}}
\end{figure}

\begin{figure}
\resizebox{8.8cm}{!}{\includegraphics{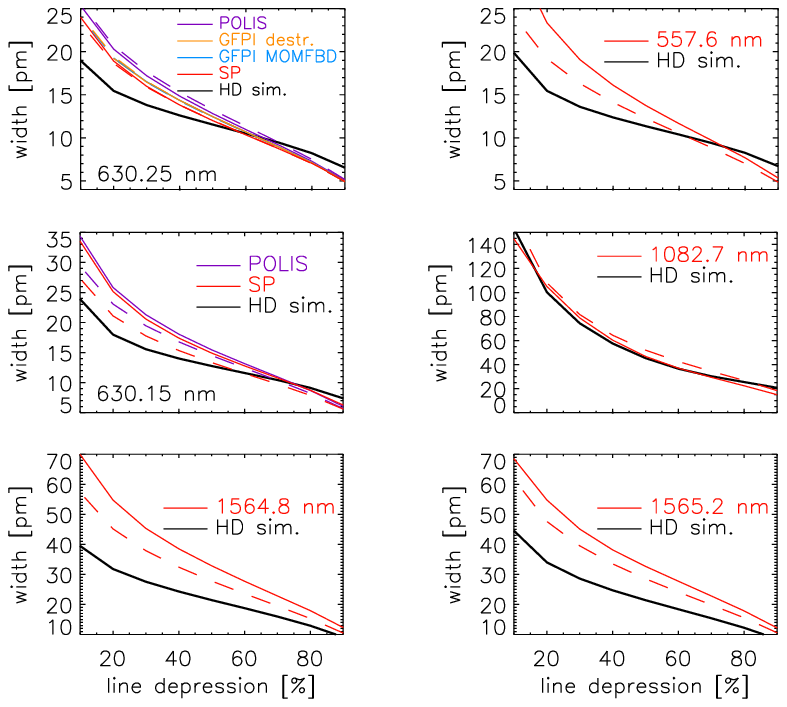}}
\caption{Average line width at different line depression levels in the
  observations (solid lines), the HD-SPAT-SPEC (dashed
    lines), and the HD-FR (solid black line).\label{bisec_length}}
\end{figure}

Figures \ref{mean_bisc_pos_rest} to \ref{bisec_length} show the average values of the line properties at different line depression levels but for the average bisector velocity of the 630.25\,nm line. The average bisector velocity of the 630.15\,nm is similar to that of 630.25\,nm in Fig.~\ref{fig_bisec}. The two near-IR lines (bottom row of Fig.~\ref{mean_bisc_pos_rest}) sample only low layers in the solar atmosphere. For the observations, the curves thus correspond to those from 10\,\% line depression up to roughly the minimal bisector velocity of the 630\,nm lines at 30\,\% line depression. The results derived from the HD-FR spectra show nearly straight bisectors for the near-IR lines. For both the 557.6\,nm and 1082.7\,nm line, the line properties close to the continuum ($< 40\,\%$ line depression) suffer from the presence of line blends (557.6\,nm) or ill-defined line depression levels (1082.7\,nm) because of the extended line wings. For the latter line, the limited extension of the simulation box is also likely to play a role. The HD-SPAT-SPEC actually gives smoother curves for these lines than the observations themselves. The average intensities at different line depression levels (Fig.~\ref{bisec_int}) provide little information in themselves because they basically only measure the line depth. The observations and the HD-SPAT-SPEC match well in most cases, with nearly identical values for instance for the 557.6\,nm or 1565.2\,nm lines. The average line widths at different line depression levels (Fig.~\ref{bisec_length}) show again that the near-IR line only sample the lower range of the atmosphere, with their curves corresponding to those of the 630\,nm lines up to the inflection point at about the 50\,\% line depression. The observations and the (degraded) HD simulation match well in the general shape of the curves here, whereas the absolute values of the line width slightly differ (cf.~the red and red-dashed curves in Fig.~\ref{bisec_length}). This mismatch in line width was already seen in the average profiles (Fig.~\ref{av_prof}). 

\subsection{Rms fluctuations}
\begin{figure}
\resizebox{8.8cm}{!}{\includegraphics{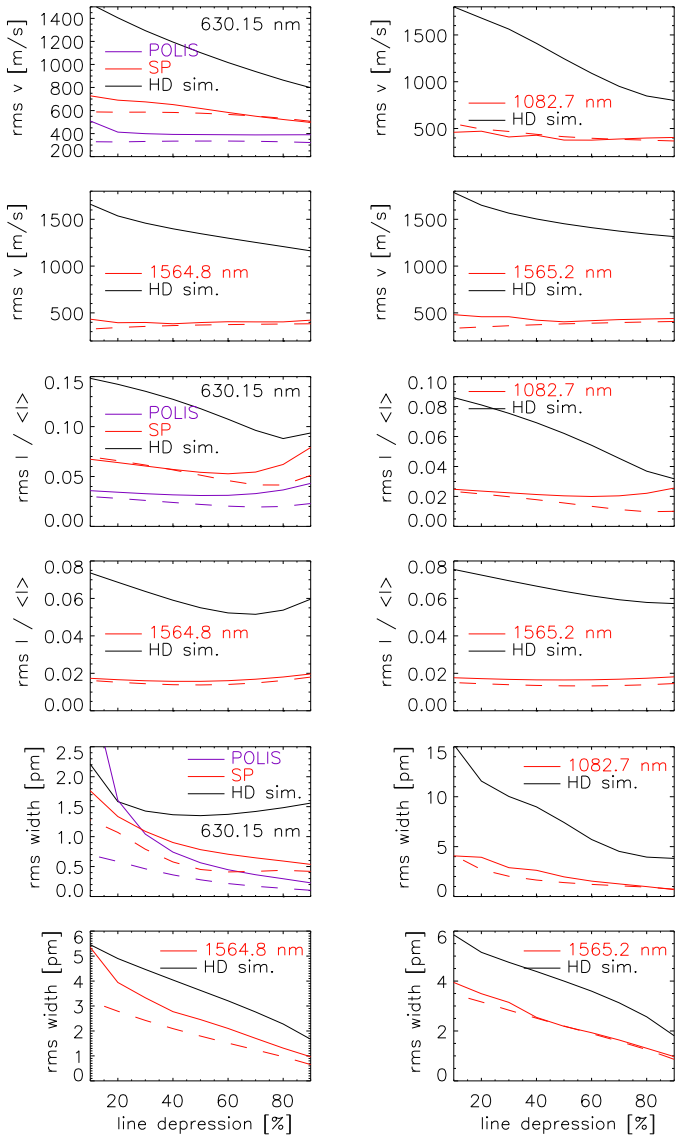}}
\caption{Rms fluctuations of the line properties at different line depression
  levels for all lines but 630.25\,nm and 557.6\,nm in the observations (solid lines), the HD-SPAT-SPEC (dashed lines), and the HD
  HD-FR (solid black line). {\em Top two rows}: velocity. {\em Middle two rows}: intensity. {\em Bottom two rows}: line width.\label{bisec_rms_rest}}
\end{figure}
The rms fluctuations of the line properties at different line depression levels for all but the 630.25\,nm and 557.6\,nm lines are shown in Fig.~\ref{bisec_rms_rest}. The curves are all to some extent similar to the corresponding ones of the 630.25\,nm line that are described in Sect.~\ref{bisec_sec}. The most noteworthy feature is that the fluctuations in the HD-SPAT-SPEC never significantly exceed the observed fluctuations in any of the parameters, and rather tend to show slightly lower values. Some parameters show a quite close match between observations and the degraded HD simulation, e.g.~the rms velocities for all lines or the intensity rms for the near-IR lines.
\section{Line parameters of 557.6\,nm \label{appb}}
\begin{figure}
\resizebox{8.8cm}{!}{\includegraphics{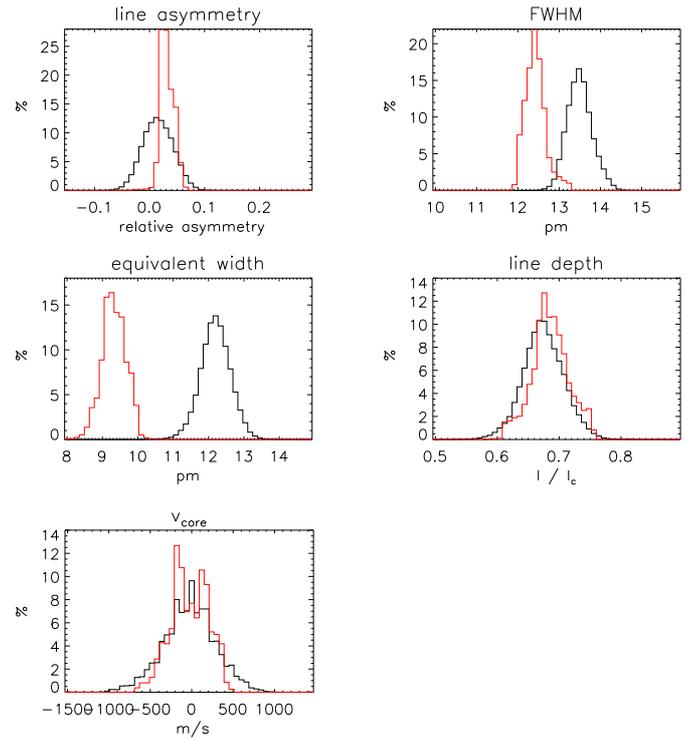}}
\caption{Line parameters of the 557.6\,nm line. Black lines:
  observations. Red lines: degraded HD simulation.\label{fig_5576}} 
\end{figure}
\begin{figure}
\resizebox{8.8cm}{!}{\includegraphics{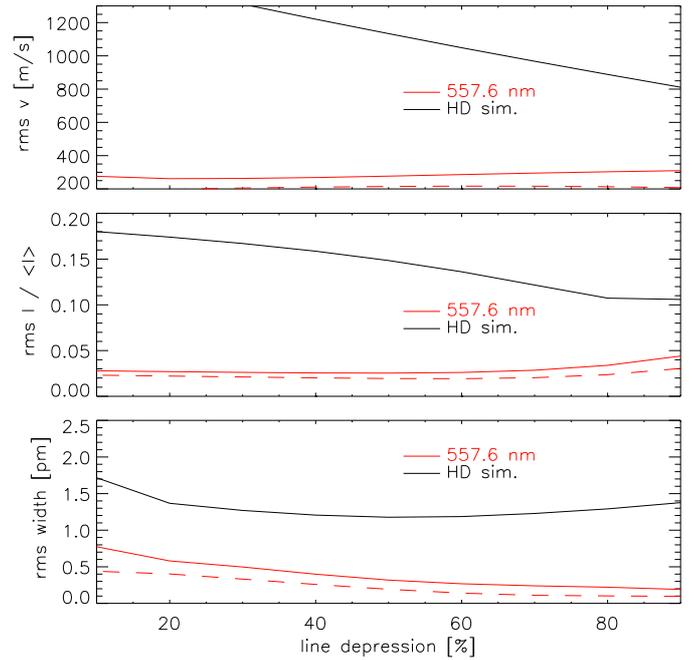}}
\caption{Rms fluctuations of the line properties at different line depression
  levels for the 557.6\,nm line in the observations (solid lines), the
  HD-SPAT-SPEC (dashed lines), and the HD-FR (solid
    black line). {\em Top}: bisector velocity. {\em Middle}: bisector intensity. {\em Bottom}: line width.\label{bisec_rms_5576}}
\end{figure}
Figure \ref{fig_5576} shows all line parameters of the \ion{Fe}{i} line at
557.6\,nm for completeness.  The 557\,nm data were intended to be used for an abundance determination and therefore taken even while the seeing conditions were rather bad. This quite limits their usefulness for deriving spatial variations of line parameters because all observed spectra already represent some large-scale spatial average. The line is the only one in the sample with a Land\'e factor of zero, i.e. not sensitive to Zeeman broadening. It thus only reacts to the thermodynamic properties of the atmosphere and not directly to the magnetic field. The line parameters (Fig.~\ref{fig_5576}) show some deviation between the observations and the HD-SPAT-SPEC, especially the FWHM and the equivalent width. This is most likely related more to the settings in the spectral synthesis, i.e.~the method for the inclusion of line broadening, than the thermodynamics themselves. The rms fluctuations of the line properties at different line depression levels (Fig.~\ref{bisec_rms_5576}) in observations and the HD-SPAT-SPEC are more similar than the histograms of line parameters. They show again lower fluctuations in the HD-SPAT-SPEC than in the observations. The 557.6\,nm line will be most important in the comparison of the thermodynamic characteristics between the field-free, purely HD and the MHD simulations of increasing average magnetic flux. In the present context, this line shows no prominent deviations from the other Zeeman-sensitive lines in its behavior.
\end{appendix}
\end{document}